\documentclass[a4paper,fleqn,usenatbib]{mnras}

\usepackage[T1]{fontenc}
\usepackage{ae,aecompl}

\usepackage{graphicx}	
\usepackage{amsmath}	
\usepackage{amssymb}	
\usepackage{mathtools}

\newcommand{\kms} {{\rm \, km \, s^{-1} }} 
\newcommand{\GAIA} {{\em GAIA}\ }
\newcommand{\au} {\, {\rm AU}}   
\newcommand{\kau} {\, {\rm kAU}} 
\newcommand{\msun} {\,M_\odot} 
\newcommand{\muasy} {\, \mu{\rm as} \, {\rm yr}^{-1} } 
\newcommand{\pc} {\, {\rm pc}} 
\newcommand{\kpc} {\, {\rm kpc}} 
\newcommand{\msecsq}{\, {\rm m \, s^{-2}}}
 
\newcommand{\vthree}{v_{\rm 3D}} 
\newcommand{\Geff}{G_{\rm eff}} 
\newcommand{\newtwo}{ }  

\title[Modified Gravity, Wide Binaries and GAIA]
 {Testing Modified Gravity Theories via Wide Binaries and GAIA}
 
\author[C. Pittordis \& W. Sutherland]{
Charalambos Pittordis,$^{1}$\thanks{E-mail: c.pittordis@qmul.ac.uk}
Will Sutherland,$^{1}$\thanks{E-mail: w.j.sutherland@qmul.ac.uk}
\\
$^{1}$ School of Physics \& Astronomy, 
 Queen Mary University of London, Mile End Road, London E1 4NS, UK.\\
}

\date{Submitted to MNRAS 29 Nov 2017; revised 11 Jun 2018; 
 accepted 12 Jun 2018}

\pubyear{2018}

\begin{document}
\label{firstpage}
\pagerange{\pageref{firstpage}--\pageref{lastpage}}
\maketitle 

\begin{abstract}
The standard $\Lambda$CDM model based on General Relativity (GR) 
 including cold dark matter (CDM) is 
 very successful at fitting cosmological observations, but recent
 non-detections of candidate dark matter (DM) particles mean that
  various modified-gravity theories remain of significant interest. The
 latter generally involve modifications to GR below a critical 
  acceleration scale $\sim 10^{-10} \msecsq $.   
 Wide-binary (WB) star systems with separations $\ga 5 \kau$ 
  provide an interesting test for 
  modified gravity, due to being in or near the low-acceleration regime  
   and presumably containing negligible DM. 
 Here, we explore the prospects for
  new observations pending from the {\em GAIA} spacecraft to  
 provide tests of GR against MOND or TeVes-like theories in a
 regime  {\newtwo  only partially explored to date}. 
  In particular, we find that 
   a histogram of (3D) binary relative velocities, relative to 
  {\newtwo equilibrium} circular velocity 
  predicted from the (2D) projected separation
  predicts a rather sharp feature in this distribution for standard gravity, 
  with an 80th (90th) percentile value close to 1.025 (1.14) 
   with rather weak dependence on the eccentricity distribution.   
  However, MOND/TeVeS theories produce a shifted distribution, with
  a significant increase in these upper percentiles.  In
  MOND-like theories {\em without} an external field effect, there
 are large shifts of order unity.  With the external field effect
 included, the shifts are considerably reduced to $\sim 0.04 - 0.08$, but
  are still potentially detectable statistically given reasonably large 
  samples and good control of contaminants.  
 In principle, followup of {\em GAIA}-selected wide binaries with ground-based 
 radial velocities accurate to $\la 0.03 \kms$ should be able to 
 produce an interesting new constraint on modified-gravity theories. 
\end{abstract}




\section{Introduction}
\label{sec:intro} 

Einstein's theory of General Relativity (GR) provides the best 
 known description of gravity on all scales. However,  
 much cosmological data \citep{Planck 2015} requires an additional cold, 
 non-baryonic \& non-visible dark matter
 (DM) component to match many observations, in addition to dark energy
  such as a cosmological constant.  
 At the present time there is no decisive direct detection of DM 
 (e.g. \citet{LUX17}); 
 this leaves an open window for possible modified-gravity theories 
  which may possibly 
  account for these effects without the inclusion of exotic DM.

The MOdified Newtonian Dynamics (MOND) is a notable theory that attempts to
 explain weak-field/non-relativistic gravitational effects without DM. This
 theory was first proposed by \citet{Milgrom 1983} to explain the flat
 rotation curves observed in most spiral galaxies without requiring DM. 
 The original MOND formulation was non-relativistic and really a
  fitting function rather than a realistic theory; 
 it has later been incorporated into relativistic theories following from
 the well-known Tensor-Vector-Scalar (TeVeS) theory proposed by
\citet{Bekenstein 2004}. 

So far, no modified-gravity theory (without DM) has been close to 
  successful in fitting the cosmic microwave background (CMB) 
 observations from {\em WMAP} and {\em Planck} \citep{Planck 2015},
 hence the $\Lambda$CDM model remains the standard model. 
But, there is a large model space for modified gravity which remains
  only partially explored, 
 and currently there does not exist any ``No-Go theorem'' demonstrating 
  that no plausible modified-gravity theory could match the CMB and
 other cosmological data in the future.  
  In the absence of either a convincing direct detection of dark matter,
  or a future general No-Go theorem, or new observations 
  excluding modified gravity at the relevant very low accelerations,  
 the situation is likely to remain unsettled; new tests which can
 discriminate between DM and modified-gravity are highly desirable.  

\text{}
\\
 In this work we consider the prospects for a test of gravity 
 in the low-acceleration regime via wide binary (WB) stellar systems;  
 previous work in this area has been done by 
 \citet{Hernandez 2011},\citet{Hernandez 2012},
 \citet{Hernandez 2014} and \citet{Matvienko 2015}; 
{\newtwo these gave hints of deviations in the direction 
  expected from MOND-like gravity, though due to the limited precision of
 current data, these hints are not yet decisive.} 

 Here {\newtwo we extend on the work of Hernandez and co-workers above, 
  but focusing on the much improved precision which will be 
  possible with GAIA data; we also explore the 
   external field effect of MOND and add new statistical tests. 
 } 

 {\newtwo Similar to \citet{Hernandez 2011}, \citet{Hernandez 2012}, 
  \citet{Hernandez 2014} and \citet{Matvienko 2015}, 
 } 
 we consider WBs with separations 
  $r \ga 3\kau$, 
 {\newtwo for which recent samples have 
 been selected by e.g. } 
 \citet{Scarpa 2017}; \citet{Coronado 2015};
 \citet{Andrews 2017}. 

 For a typical stellar WB 
  with a separation $r \sim 7000 \, {\rm AU}$, 
 and masses $M \sim 1\,M_{\odot}$, the acceleration is 
  $a\sim 10^{-10} \msecsq$ which is comparable to the 
 critical acceleration constant $a_0$ in MOND-like theories, and also    
  similar to the local gravitational acceleration due to  
  our Milky Way galaxy. 
\\
\text{}
\\
The formation mechanism of WBs is not well known, but may well 
 result {\newtwo from } captures during evaporation of star-forming clusters; 
 the key point for the present purposes 
 is that WBs {\newtwo are not expected to } contain any significant amount
 of DM, so their distribution in orbital parameters should follow
 GR/Newtonian predictions apart from perturbations from Galactic tides, 
  giant molecular clouds and passing stellar fly-bys.  These
 perturbations are significant, but disrupted binaries should 
 separate out to many-parsec separations 
   on a timescale $\sim 10$ Myr which is much shorter than
 the age of the Galaxy;  thus there should be a reasonably 
 clear distinction between currently-bound  
 and disrupted wide binaries. 

Previous work of \citet{Yoo 2003} \& \citet{Quinn 2009}
  used a WB sample selected by \citet{Chaname 2004}, 
  and examined the distribution of angular separations.  The observed
 distribution is consistent with a power-law; tidal disruption 
 by an external perturbing source such as Galactic \&
 disk tides, giant molecular clouds or massive compact halo objects (MACHOs)
  would  preferentially disrupt the widest binaries, 
 inducing a break in this power-law at large separations. 
 An upper limit on such a break can bound the abundance
 of massive perturbers, placing upper limits on the abundance of  
  dark-halo MACHOs of mass $\ga 100 \msun$. 
\\
The main observational challenge with WBs is that their orbital periods are 
 extremely long (and accelerations and velocity differences 
  very small), so realistic observations
 provide only a instantaneous snapshot of position and velocity differences,
  and individual orbit solutions are not possible.  However, statistical
 distributions of velocity differences 
 for a large sample of wide binary systems can still provide 
 an interesting constraint, as we explore below. 

In this paper we numerically compute the observables for samples of 
 simulated WB systems with various assumed gravity models, including
  GR/Newton and several MoND/TeVes models. We 
 predict their \textit{velocity ratio vs projected separation} 
 distribution that would be derived from on-going observations 
 with {\em GAIA}, and future observations with high-precision ground-based
  radial velocity measurements. The plan of the paper is as follows: 
 in Section~\ref{sec:wb-gaia} 
  we provide an overview of the issues, define our
   notation and recap some standard results for Newtonian orbits. 
 In Section~\ref{sec:mg} we review some selected 
  theories of modified gravity as they pertain to the next sections.  
 In Section~\ref{sec:orbs} we produce simulations of WB orbits for various
   gravity theories, and we produce forecasts of observables. 
 In Section~\ref{sec:obs} we discuss some additional observational issues, and  
 in Section~\ref{sec:conc} we summarise the conclusions.  

\section{Wide Binaries and GAIA} 
\label{sec:wb-gaia} 

Here we define ``wide binaries'' as those with orbital 
 accelerations comparable to the MOND acceleration parameter $a_0$, 
 commonly defined as $a_0 \simeq 1.2 \times 10^{-10} \msecsq $,
  or separations above $\sim 7 \kau$ for Solar-mass binaries.   
 Wide binaries have received relatively little attention in the
 past for several reasons: their orbital periods ($\sim$ Myr) are so long that
  no deviations from linear motion are detectable 
   on a realistic timescale: 
  thus full orbit solutions are impossible, and any reasonable
   observing programme gives us only a snapshot of some subset of the six 
  phase-space parameters (three relative positions and three
  relative velocities).   Also, their relative velocities
 of order $\sim 0.3 \kms$ translate to expected proper motion {\em differences} 
 of order 0.6 milliarcsec (mas) per year at an example distance of 100 pc, 
  which is near the limits of achievable $1\sigma$ measurement precision 
  in the pre-\GAIA era. 
  Finally, uncertainties in available parallax distances translate
    to significant stellar mass uncertainties which further 
   blurs any possible constraints. 
 Thus, wide binaries can be {\em selected} fairly robustly with pre-\GAIA data 
  as from Hipparcos \citep{Lepine 2007}, the SlowPOKES search from SDSS
   \citep{slowpokes}, and \citet{Tokovinin 2016}, using
 proper motions to reject most chance-projection candidates or unbound fly-by
  pairs;  but the relative velocity precision is currently 
   not sufficient to use the resulting binaries for dynamical tests. 

 Also, there is a notable gap in previous wide binary catalogues: Hipparcos 
  parallaxes are generally limited to magnitude $V \la 10$, while SDSS imaging
  saturates for stars at $V \la 14$, which leaves the magnitude
   range $10 < V < 14$ rather less explored for wide binaries. 
 This magnitude range includes millions of stars, with potentially
  $\sim 0.5$ million closer than $\approx 200 \pc$, and many 
   thousand wide binaries (see Section~\ref{sec:num}) 
  which are bright enough to follow-up with high-quality 
  ground-based spectroscopy. 

 The \GAIA spacecraft \citep{gaia-miss} 
  will dramatically transform this situation:
  firstly, its proper motion precision after the baseline 
   5-year mission is predicted to be around
  15 microarcsec per year ($\muasy$)
   at magnitude $G \simeq 15$ \citep{gaia-miss}; 
  multiplying by $\sqrt{2}$ for the differential
  motion between two stars translates to a very small 
  transverse velocity uncertainty of $\sim 0.01 \kms$ at our 
  example $100 \pc$ distance,
  which is much smaller than expected binary orbital velocities at
  separations $\sim 10 \kau$.   
  Secondly, \GAIA will provide high-precision parallax distances 
  (better than 1\% for the above parameters) and 
  hence precise luminosities, while
   metallicities can readily be obtained either from the GAIA spectra or 
  from the ground for these bright stars. 
   Using a mass-luminosity-metallicity relation, 
 for main-sequence stars we can then infer masses for
  both components of wide binaries; uncertainties such as age may limit this
   to perhaps 5 percent mass precision, 
  but this is essentially good enough for the following purposes. 

 The \GAIA distance precision (e.g. 0.3 percent at $100 \pc$) 
 is  usually not quite good enough to resolve the line-of-sight 
 separation of a typical wide binary at $r \sim 3 - 20 \kau$ 
  (though for very nearby and wide systems 
  at $\sim 20 \pc$, resolving the line-of-sight separation 
  should be possible);  
  however it is good enough to weed out the vast majority of
  chance-projection interlopers: observed star pairs with small projected
  separation $\la 20 \kau$, 3-D velocity difference $< 1 \kms$
  and line-of-sight separation $\la 0.5 \pc$ are highly likely
  to be either true bound binaries, or unbound but physically associated 
  pairs with a common origin, since unrelated chance-flyby pairs with such 
  small differences should be very rare (see later).   

  For radial velocities,  the \GAIA precision
    is not good enough, but modern high-stability ground-based
  spectrographs can reach absolute accuracy $\sim 0.02 \kms$ 
  (probably limited by systematics, see Section~\ref{sec:obs} later).   

  Thus, with \GAIA plus high-accuracy radial velocity followup, 
  we can get fairly precise measurements of 5 of the 6 phase-space
  differences for wide binaries 
  (subject to perspective-rotation effects, discussed
  later), with the relative-velocity precision of order 10 percent; 
    this turns out to be enough to get potentially
 interesting tests of gravity in the low-acceleration regime 
  where any modified-gravity effects should start to 
  become significant.    

\subsection{External perturbations on wide binaries} 

{\newtwo Wide binaries are weakly bound and thus are significantly
 sensitive to perturbations from either fly-by encounters
 with passing stars, from giant molecular clouds, or from Galactic
 tidal effects. Tidal effects are expected to disrupt
 binaries beyond the Jacobi radius 
  around $1.7 \pc$ or $350 \kau$ for a typical binary; 
  this is over an order 
 of magnitude larger than the separations considered below, 
  but tidal effects may be non-negligible at smaller scales.  
  A numerical simulation of these effects has been
 made by \citet{jiang-tremaine} (JT10), with the following main conclusions.

 The survival probability for a wide binary over 10~Gyr is a
  declining function of semi-major axis, with estimated 50 percent
 survival probability occurring at a separation of around $30 \kau$. 
  Binaries which become unbound do not always separate completely,  
  but can remain within 10 parsec separation for many Gyr after
  unbinding. The histogram of separations for binaries evolved
 over 10 Gyr shows a minimum at $\sim 3 \, r_J $ or $5 \pc$, then a secondary
  maximum at $\sim 10 \pc$. 
 Also, Figure~3 of JT10 shows that the distribution 
 of projected separations at $r \la 10 \kau$ largely follows the initial
 distribution, which is promising for the tests below. This suggests
  that external perturbations may significantly randomise the 
 eccentricity distribution of binaries at $\sim 10 \kau$, but we see below
 that our results are relatively insensitive to this poorly-known
 eccentricity distribution.   
 The  RMS (line of sight) velocity difference of the simulated 
   binaries in JT10 closely follows the expected Keplerian 
 falloff $\propto r_p^{-0.5}$ out to 
 projected separations $\sim 0.3 \, r_J$ or $100 \kau$ (as shown 
  in Figure~7 of JT10). 
 
  Since our main focus below is on binaries of 
  present-day projected separation between $3$ to $20 \kau$, we expect that
  external perturbations are not a major source of uncertainty
 in this range, though further numerical work would be desirable to
  quantify this more precisely.  
} 

\subsection{Distribution of velocity differences} 
\label{sec:u3d} 

We start here considering an idealised case assuming a wide binary where  
  both masses and all six relative separation
 and velocity components are reasonably well measured;
  then consider practical deviations from this later. 

If we have a candidate 
 binary of estimated masses $M_1, M_2$ at (3D) separation $r$, 
 we define a convenient dimensionless parameter 
 $u_{3D} \equiv \vthree / v_C(r)$ where $\vthree$ 
 is the magnitude of the instantaneous 
  (3D) velocity difference, 
  and $v_C(r)$ is the velocity for a circular Newtonian orbit 
  at the current separation $r$  
 (note, {\em not} the unknown semi-major-axis $a$). 
  Clearly $v_C(r) = \left[ G(M_1+M_2)/r \right]^{0.5}$. 

In terms of the semi-major axis $a$ and eccentricity $e$, 
  we then have
\begin{equation}
  u_{3D} = \sqrt{2 - r/a}  \ , 
\end{equation} with the well-known result 
 that $u_{3D} < \sqrt{2}$ for any bound orbit.  In general for any bound
 binary with eccentricity $e < 1$, we have $1 - e \le u_{3D}^2 \le 1+e$.  
 Considering the probability distribution for $u_{3D}$ for a large
 sample of binaries observed at a random time (i.e. now), 
  it turns out that values of $u \ga 1.2$ are
 rather uncommon, since low-$e$ binaries never exceed this value, 
 while high-$e$ binaries do so, but only for a rather 
 small fraction of time around orbit pericenter.  

For an assumed eccentricity $e$ and  an arbitrary threshold value, $u_{th}$, 
 we can readily compute  
  the fraction of time over which the instantaneous 
  value of $u_{3D}$ exceeds a chosen threshold $u_{th}$, as follows.  
 In terms of true anomaly (angle from pericenter) $\theta$, 
  we have 
 \[ 
  u_{3D}^2 = 2 - \frac{1-e^2}{1 + e \cos \theta} 
\] 
 Rearranging for a chosen threshold value $u = u_{th}$, we find in the case
  $1-e < u_{th}^2 < 1+e$ then 
   $u$ crosses $u_{th}$ twice per orbit, at $\theta$ given by  
\begin{equation} 
\label{eq:thth}
 \cos \theta_{th} = \frac{1 + e^2 - u_{th}^2}{e (u_{th}^2 - 2) } \ . 
\end{equation} 
The corresponding eccentric anomaly $E_{th}$ is 
\begin{equation} 
  E_{th} = 2 \arctan \left( \sqrt{ \frac{1-e}{1+e}} \tan(\theta_{th}/2) \right) 
\end{equation} 
 and the mean anomaly $M_{th}$ follows from Kepler's equation
\begin{equation} 
\label{eq:mth} 
 M_{th} = E_{th} - e \sin E_{th} \ , 
\end{equation} 
 where $\theta_{th}, E_{th}$ and $M_{th}$ each 
  have two solutions of opposite sign. 
 Since mean anomaly is just time rescaled to $2 \pi$ per orbit, the 
  fraction of time for which a Kepler orbit of given eccentricity 
  $e$ exceeds a chosen threshold value $u_{th}$ is then simply 
\begin{equation} 
  P( u > u_{th} \, | \, e) = M_{th} / \pi \ ,  
 \label{eq:pu} 
\end{equation}  
 which is given by successive substitutions into Eqs.~(\ref{eq:thth} -- 
 \ref{eq:pu}). 

(Note in the special case $u_{th} = 1$, this occurs on the minor axis 
 where $\cos \theta_{th} = -e$,  
 and it is easily seen that $P(u > 1) = 1/2 - e / \pi$, which decreases 
  linearly from 1/2 at small $e$ to 0.182 as $e \rightarrow 1$). 

\begin{figure}
\hspace*{-1cm}\includegraphics[width=10cm]{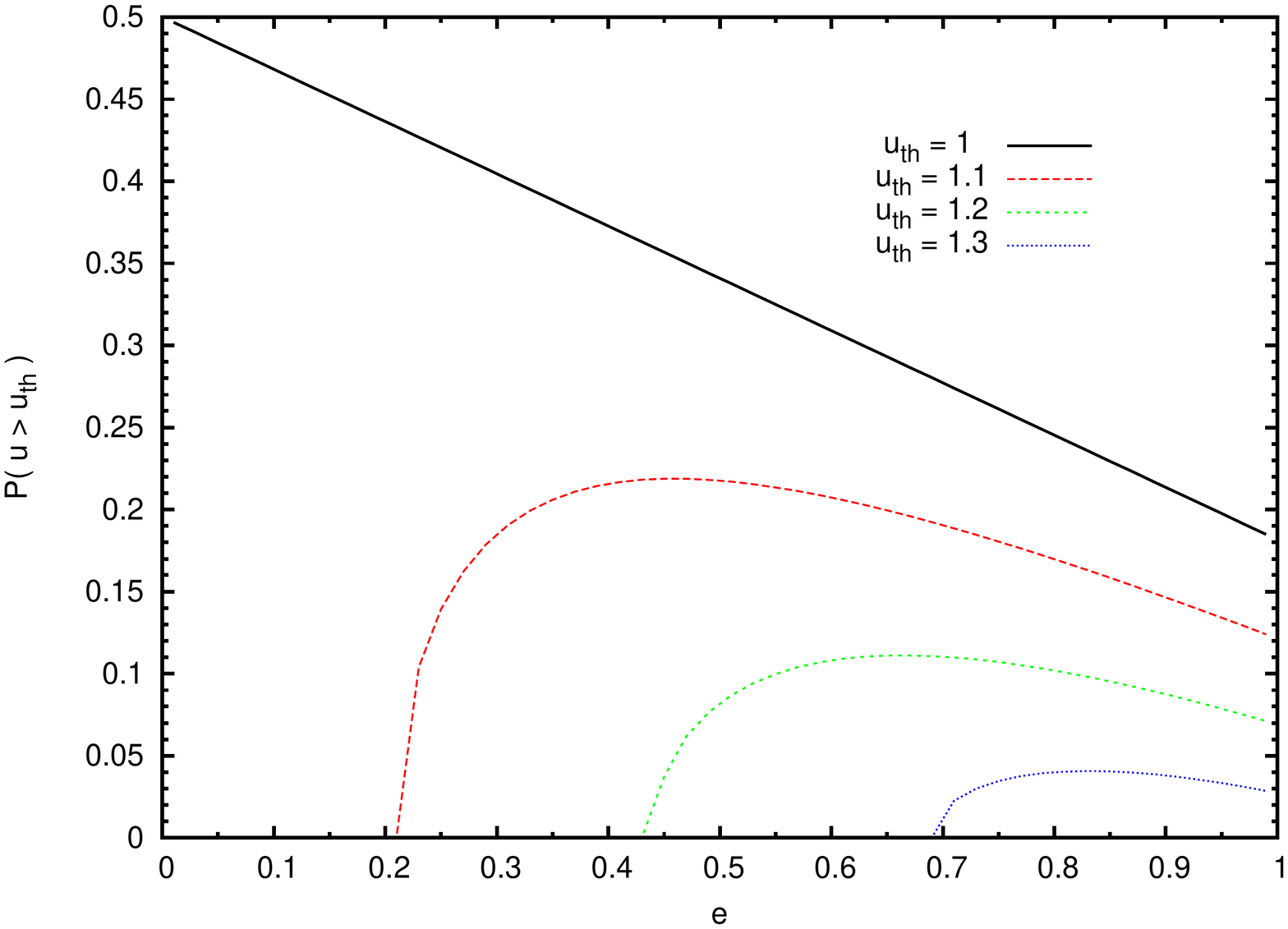} 
\vspace*{-8mm} 
\caption{Fraction of time for which a Kepler orbit exceeds 
  a selected velocity ratio $u_{th}$ as a function of eccentricity $e$,
  for values of $u_{th} = 1.0,$ 
 1.1, 1.2, 1.3 (top to bottom). \label{fig:uvse} }
\vspace{2mm} 
\hspace*{-1cm}\includegraphics[width=10cm]{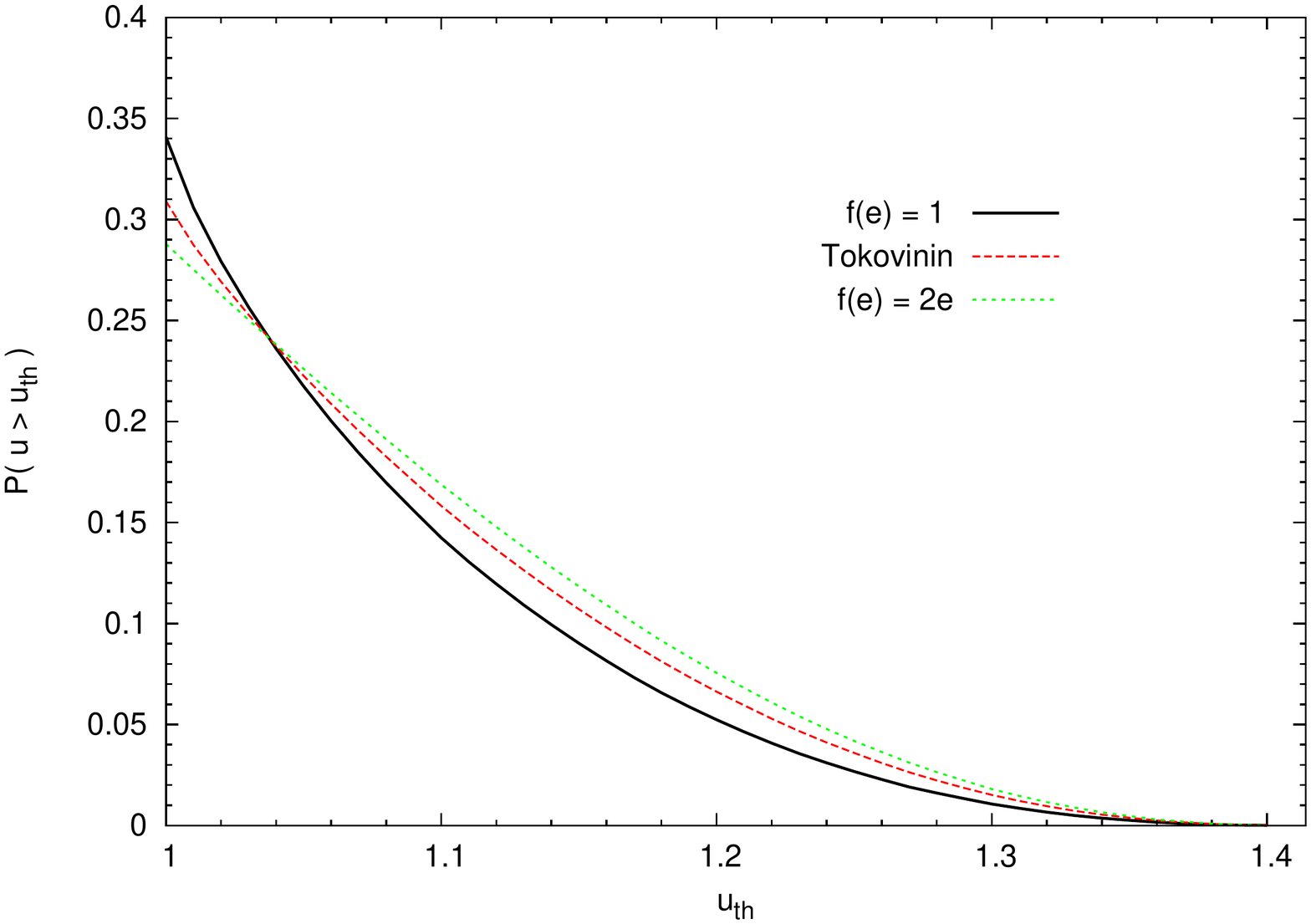} 
\vspace*{-8mm} 
\caption{Fraction of all orbits (at a random time) which exceed a
  given velocity ratio $u_{th}$ on the abscissa, 
  for several chosen distribution functions of eccentricity: 
  solid curve for $f(e) = 1$, dashed curve for Tokovinin distribution,
  and dotted curve for $f(e) = 2\,e$. 
  \label{fig:uedist}}
\end{figure} 

The resulting probability is shown as a function of $e$ for
  several example thresholds $u_{th} = 1.0, 1.1, 1.2, 1.3$ 
 in Figure~\ref{fig:uvse}. 
  The result (for the case $u_{th}>1$) is that  
  this probability is zero for $e < u_{th}^2 - 1$, then
  rises rather quickly to a maximum then slowly declines towards $e = 1$. 
 The maximum probability is 0.219, 0.111 and 0.041 respectively
  for $u_{th} = 1.1, 1.2, 1.3$,  therefore a notable feature
  here is that  
 the fraction of bound binaries with $u > 1.2$ (at a random observing time) 
 cannot exceed 11.1 percent (the case for a distribution of $e$ sharply
 peaked around $e \approx 0.67$), 
  while for a realistic spread of $e$ values,
  the fraction must be smaller than 11.1 percent. 


 To get the predicted distribution of $u$ for a large sample
 of binaries with an assumed 
  distribution function of eccentricities $f(e)$, 
  we can simply integrate the above function in Eq.~(\ref{eq:pu}) 
  weighted by the assumed $f(e)$. The result of this is shown 
  in Figure~\ref{fig:uedist} 
 for three selected distributions of $e$: firstly a uniform distribution 
 $f(e) = 1$, secondly a Tokovinin distribution $f(e) = 0.4 + 1.2e$, 
 and thirdly a ``dynamical'' distribution $f(e) = 2\,e$. 
  Due to the shape of $P(u > u_{th} | e)$ above, the result turns
 out to be only weakly sensitive to the eccentricity
  distribution, and there is always a steeply-falling tail at $u \ga 1.2$.  
 Taking the Tokovinin distribution as an intermediate case, 
  we find that (at a random time) 
  30.9 percent of binaries have $u_{3D} > 1$, while 15.8 percent
  have $u > 1.1 $, 6.6 percent have
  $u > 1.2$, and only 1.5 percent have $u > 1.3$. 
 The 80th and 90th percentiles are at $u = 1.065$ and $1.158$ 
  respectively.  
 These percentages change only marginally for the flat or dynamical
  eccentricity distributions.    

Therefore, in an idealised case of a moderately large but plausible 
  sample of candidate wide binaries (e.g. few hundred to few thousand) 
  all with $u_{3D}$ measurements, 
 standard gravity predicts that a histogram of $u_{3D}$  
  should exhibit a smoothly rising distribution 
  at $0 < u < 1$ followed by a 
  rather steep decline between $u \sim 1.1$ and $1.3$; and
   the location of this ``ramp'' feature around the 80th to 90th percentiles 
   is only weakly sensitive to the poorly known distribution of $e$.  
  The inevitable contamination from unbound
  pairs is expected to show a relatively flat or moderately
 rising distribution of $u_{3D}$ at $u_{3D} > \sqrt{2}$; 
  we may need to model this contamination and subtract an estimated
  number of contaminants scattered to $u_{3D} < \sqrt{2}$,  
 but as long as unbound chance pairs do not dominate the sample,
  the 80th/90th percentiles for bound binaries should be 
  statistically correctable for contamination. 

\subsection{Projection to 2D separations} 
 The above was assuming an idealised case where 
  all six components of separation and 
  relative velocities are available: 
  but in practice, $u_{3D}$ is not directly measurable due to 
   the uncertain line-of-sight component of the separation vector; 
 however, the 3D relative velocity and 
  the 2D projected separation {\em are} accurately measured, so 
  we can make do with $u_{2D}$, defined as the ratio of
 3D relative velocity to the circular velocity calculated at
 the observed 2D projected separation.  This is then given
 by 
 \[ 
  u_{2D} = u_{3D} \sqrt{\sin \beta} 
\] 
 where $\beta$ is the unknown angle 
  between the current binary separation vector 
  and the line-of-sight;  for random aligments, the median 
 value of $\sin \beta$ is $\sqrt{3}/2$, so the median of the 
  $\sqrt{\sin \beta}$ factor is 0.931, only slightly less than 1. 
  Therefore, the effect of convolution with random aligment angles 
  is to shift the distribution function of $u_{2D}$ to somewhat
  smaller values compared to $u_{3D}$, but it does not erase the 
   steep decline in the distribution. 
  The quantitative effects of this projection to 2D projected separations 
   are included below 
  in Section~\ref{sec:orbs}, using numerical simulations.  

 We note that a uniform distribution in $\cos \beta$ gives the distribution
   of projected separation $r_p$ at a given 3D separation $r$.   
  The inverse question, of the posterior probability distribution of 
   $r$ at a known $r_p$, is not quite the 
  same question, and in this case the result depends
  on the intrinsic frequency distribution of $r$;   
  however, assuming a reasonably smooth distribution in $\log r$ this 
   distinction is relatively unimportant. 
 
\subsection{Perspective effects} 
\label{sec:perspec} 
 It was shown by \citet{shaya-olling} (hereafter SO11)   
  that there are several effects implying that
  wide binary velocity differences are not simply given by subtracting
 the measured proper motions and radial velocities; we
 refer to these collectively as perspective effects, since they are
 mostly related to the system barycentre motion (relative
 to the Sun) causing a time-dependent perspective on the system 
  from our location.  

 SO11 gave a derivation of these effects to first order
 in binary separation angle $\alpha = (\Delta \ell, \Delta b)$ 
  with the key results given in their Eqs. 29, 30. 
We find that is somewhat more intuitive 
 to re-arrange SO11 Eq.~29 into the form 
\begin{equation} 
\label{eq:so29} 
 \Delta{\bmu} = -\bmu \, \frac{\Delta d}{d} 
 + \Delta \ell \sin b \begin{pmatrix} \mu_b \\ -\mu_\ell \end{pmatrix} 
 - \frac{v_r}{d} \begin{pmatrix} \Delta \ell \cos b \\ \Delta b 
  \end{pmatrix} \ . 
\end{equation} 
 which gives the apparent proper motion difference for a 
  hypothetical ``static'' binary;  
 where $(\ell, b)$ are Galactic coordinates, $\bmu = (\mu_\ell, \mu_b)$ 
  is the proper motion 2-vector and $\mu_\ell \equiv (d\ell/dt) \cos b$ 
    includes the $\cos b$ factor;  
  $d$ is distance and $v_r$ is radial
   velocity of the barycentre, and $\Delta$'s denote differences
  between the two components of the binary. 
 
 These three terms each have an intuitive geometrical explanation: 
 the first term on the RHS corresponds to the closer component appearing
  to overtake the more distant component 
 in the direction parallel to the proper motion.  

 The second term is now seen as a pure coordinate-curvature effect: 
 if we consider tangent-plane coordinates at the barycentre, 
 constant-$b$ lines appear as conic sections, which are locally
  equivalent to circular arcs with a radius of curvature 
   of $\cot b$. The component of binary 
  separation in the $\ell$ direction is $\approx \Delta \ell \, \cos b$, 
 hence the constant-$b$ curves through the primary and secondary
  have a relative rotation angle 
 in the tangent plane by $\approx \Delta \ell \cos b / \cot b = 
  \Delta \ell \sin b$ .
 The second term above therefore is equivalent to the difference
  between the barycentre proper motion vector, and a copy of itself 
  rotated by this (small) angle. 

The third term is simply an apparent contraction/expansion of the binary
  angular separation at a fractional rate of $-v_r/d$. 

 Also,  Eq. 30 of SO11 may be written as 
 \begin{equation}
\label{eq:so30} 
 \Delta v_r = \begin{pmatrix} d \mu_l \\ d \mu_b \end{pmatrix} \cdot 
  \begin{pmatrix} \Delta \ell \cos b \\ \Delta b \end{pmatrix} 
\end{equation}   
 which is the scalar product of the tangential velocity vector 
  with the binary separation angle. 
 This may be combined with $d \times$ the first term in Eq.~\ref{eq:so29}, 
 in which case the resultant 3D velocity corresponds to a rotation
  of the 3D binary separation vector 
  at an angular speed $\mu$ around the line 
  perpendicular to both the line of sight and the 
  proper motion vector, i.e. a ``perspective rotation''.  
  This perspective rotation 
  effect is important; the effect on $\Delta v_r$ in Eq.~\ref{eq:so30} 
  is calculable given the known angular offset and proper motion, 
  but the effect on transverse velocity in Eq.~\ref{eq:so29} 
  is proportional to $\Delta d / d $, and $\Delta d$ is generally
  not measurable to useful precision even with \GAIA data. 

 The above analysis based on SO11
   is valid up to terms first-order in 
  binary separation angle. These are definitely important, since for
 typical values of system barycentre motion $v_{sys} \sim 50 \kms$, 
  angular separation $\alpha  
   \sim 20 \kau$ / 100 pc  $\sim 0.001 \, \text{rad}$, 
 and $\Delta d / d \sim 0.001$,  terms of order $v_{sys} \alpha$ 
  are $\sim 0.05 \kms$, of order 20 percent of the binary relative velocity 
   $\sim 0.25 \kms$; {\newtwo this is modest but not negligible.}   
 However, this may be constrained e.g. by rejecting a tail of binaries
  with higher transverse velocities; it is also helpful that 
  the effect (in velocity units) decreases with distance for
  fixed binary separation.  

 We note that terms of order $v_{sys} \alpha^2$ are generally 
  negligible except for very nearby or extremely wide 
   binaries (few-degree separations), 
  so the first-order treatment given by \cite{shaya-olling} is adequate  
  except for very nearby or {\newtwo extremely} wide binaries.   


\section{Modified gravity models} 
\label{sec:mg} 

In this section we review some of the various modified-gravity
 scenarios studied in the literature as 
 possible alternatives to dark matter; these are then applied to 
 simulated orbits of wide binaries in the following 
 Section~\ref{sec:orbs} 

\subsection{MOND}
\label{sec:mond} 

 The phenomenology known as Modified Newtonian Dynamics (MOND) 
 was originally introduced in the 1980's by \citet{Milgrom 1983}, 
  and has led to many variants and refinements later (see \citep{Famaey 2012} 
 for a comprehensive review). 
 The original motivation for MOND was to modify Newton's second law $F_N = ma$ 
  in order to attempt to account for observed 
  effects such as flat rotation curves of spiral galaxies 
  without the need for DM. 
 The modification to Newton's second law is made by introducing 
 a critical acceleration constant, $a_0$, and a free function $\mu(x)$, 
  where the dimensionless 
  $x \equiv a / a_0$ is the ratio of acceleration to $a_0$, such that   
\begin{equation}
    a_N = \mu\bigg(\frac{a_{M}}{a_0}\bigg)a_{M}.
	\label{eq:MOND} 
\end{equation}
 where $a_N$ is the GR/Newtonian acceleration, 
   $a_0 \approx 1.2 \times 10^{-10} \msecsq$ is the 
 acceleration constant and $a_{M}$ is the acceleration predicted by MOND.  
  (In general this requires numerical solution for $a_M$ given $a_N$ 
  and a chosen function $\mu$). 

 The dimensionless interpolating function $\mu(x)$  is arbitrary, 
   but is required to have  
  $\mu(x) \rightarrow 1$ for $x \gg 1$ to satisfy Solar-system
 constraints,  and $\mu(x) \sim x$ at $x \ll 1$ in order to produce 
  flat galaxy rotation curves at large radii ($a_N \ll a_0$),
  and $\mu(x)$ should be monotonically increasing between these limiting
   cases.  

  Many possible functions can be chosen given these constraints:  
  two common choices are 
 $\mu(x)= (1+\frac{1}{x})^{-1}$, known as the `Simple' interpolating 
 function, or $\mu(x)= x(1 + x^2)^{-1/2}$ known as 
 the `Standard' interpolating function, where $x \equiv {a_{M}}/ {a_o}$. 

 In the ``deep MOND'' regime, $x \ll 1$,  
  the orbital velocity  tends to remain constant, and is given by 
\begin{equation}
    v_{M} \approx (GMa_0)^{1/4}. 
	\label{eq:Mu=1}
\end{equation} 
\\
The foundation of the MOND theory begins with a non-linear 
 Poisson equation, given by:

\begin{equation}
    \nabla \cdot \left(\mu \bigg( \frac{|\nabla \Phi_{M}|}{a_0} \bigg) 
  \nabla \Phi_{M} \right)= 4\pi G \rho = \nabla \Phi_{N}
	\label{eq:non-poisson}
\end{equation}
\\
where $\Phi_{M}$ is the MOND potential, 
 $-\nabla \Phi_{M} = \ddot{r}_{M} = g_{M}$, 
 can be obtained from taking the Euler-Lagrange 
 equation of the AQUAdratic Lagrangian (AQUAL) theory of MOND, given by:

\begin{equation}
    S_{{AQUAL}} = \int d^3x \, 
  \left[ \rho \Phi_{N} + \frac{a_0^2}{8 \pi G}\mu 
 \bigg( \frac{(\nabla \Phi_{M})^2}{a_0^2} \bigg)\right]
\label{eq:aqual}
\end{equation}

From applying equation~(\ref{eq:non-poisson}), one obtains 
  equation~(\ref{eq:MOND}) \citep{BekensteinMilgrom 1984}.
\\
\subsection{TeVeS} 
\label{sec:teves} 
The original MOND theory is non-relativistic and hence can only 
 represent an approximation to some more fundamental theory in the
 low-velocity limit.   
A relativistic counterpart to the MOND theory is given by 
 the TeVeS theory invented by \citet{Bekenstein 2004}. 
 The construction of the Lagrangian of TeVeS employs a unit vector field,
  a dynamical \& non-dynamical scalar field, a free function and a 
 non-Einsteinian metric tensor, \textit{effective or physical metric} 
 in order to reproduce the MOND dynamics in the non-relativistic limits. 
 The TeVeS action is expressed as:
 
\begin{equation}
    S_{_{TeVeS}}=\int d^4x \, (\mathcal{L}_{g}+\mathcal{L}_{s}+\mathcal{L}_{v}) +
 \mathcal{L}_{matter}  
	\label{eq:s-teves}
\end{equation}

 where $\mathcal{L}_{g}$ is the Einstein-Hilbert action  
\begin{equation}
    S_{EH}=\frac{1}{16 \pi G} \int d^4x \, 
  \sqrt{-g}(R-2\Lambda) + \mathcal{L}_{m}
        \label{eq:E-H action}
\end{equation}
 and $\mathcal{L}_{m}$ 
 is the matter field within the theory. 
 Above $\mathcal{L}_{s}$ and $\mathcal{L}_{v}$ are the TeVeS 
 scalar and vector field lagrangians, which are expressed as:

\begin{align*}
	 \mathcal{L}_{s} = -\frac{1}{2}\bigg[\sigma^2 h^{\mu \nu}\partial_{\mu}\phi \partial_{\nu}\phi + \frac{1}{2}\frac{G}{l^2}\sigma^4 F(kG\sigma^2)\bigg]\sqrt{-g} \\
     h^{\mu \nu}= g^{\mu \nu}-U^{\mu}U^{\nu} \\\\
      \mathcal{L}_{v}= -\frac{K}{32G\pi}\bigg[ g^{\alpha\beta}g^{\mu\nu}(B_{\alpha\beta}B_{\mu\nu})+ \frac{2\lambda}{K}(g^{\mu\nu}U_{\mu}U_{\nu}-1) \bigg]\sqrt{-g}\\
      B_{\alpha\beta}=\partial_{\alpha}U_{\beta}-\partial_{\beta}U_{\alpha}
	\label{eq:\lambda = 1}
\end{align*}

where $k$ is a dimensionless constant, $K=\frac{k}{2\pi}$, $l$ is a  
 constant length, $\lambda$ is the coupling factor and 
 $\sigma$ is the coefficient responsible for the kinetic terms; 
 see \citet{Bekenstein 2004} for a more explicit definition for these terms.
\\
Taking the action principle and varying the action 
 with respect to its tensor, vector and scalar parts,
 Bekenstein derives the field equation for TeVeS, expressed as: 

\begin{equation}
G_{\mu \nu} = \frac{8\pi G}{c^2}(T_{\mu \nu}+(1-e^{-4\phi})U^\alpha 
  T_{\alpha}(_{\mu} U_\nu) + \tau_{\mu \nu}  )+ \Theta_{\mu \nu} \, 
\end{equation}
\\
where the terms $\tau_{\mu \nu}$ \& $\Theta_{\mu \nu}$ are expressed as

\begin{gather*}
\tau_{\mu \nu} = \sigma^2(\partial_{\mu} \phi \partial_{\nu} \phi 
 -\frac{1}{2}g^{\alpha \beta}\partial_{\alpha} \phi \partial_{\beta} \phi g_{\mu \nu}\\\
 -U^{\alpha}\partial_{\alpha} \phi (U(_{\mu}\partial_{\nu}\phi)  
 - \frac{1}{2}U^{\beta}\partial_{\nu}\phi g_{\mu \nu}) ) \\
- \frac{1}{4}Gl^{-2}\sigma^4 F(kG\sigma^2)g_{\mu \nu}
\end{gather*}
 
\begin{align*}
\Theta_{\mu \nu} = k(g^{\alpha \beta}F_{\alpha \mu}F_{\beta \nu} -
  \frac{1}{4}F_{\alpha \beta}F^{\alpha \beta}g_{\mu \nu})- \lambda U_{\mu}U_{\nu}
\end{align*}
\subsubsection{MOND approximation}
Deriving the weak-field limit in TeVeS, one can reproduce the MONDian 
 dynamics via deriving the '\textit{physical metric}' 
 Equation~(\ref{eq:metric_TeVeS}) and computing the geodesics, 
 Equation~(\ref{eq:geod_TeVeS}), \citep{Bekenstein 2004}:

\begin{equation}
\tilde{g}_{\mu \nu} = e^{2\phi}g_{\mu \nu} - 2U^{\mu}U^{\nu} \,\sinh(2\phi)
	\label{eq:metric_TeVeS}
\end{equation}
 
\begin{equation}
\tilde{\Gamma}^{\lambda}_{\mu \nu} = e^{-2\phi}\Gamma^{\lambda}_{\mu \nu}
	\label{eq:geod_TeVeS}
\end{equation}

By solving the TeVeS field equation in the weak-field limit using only the 
 leading order terms $h^2_{00}$ \& $h^2_{11}$ derived from the TeVeS physical 
 metric, we obtain the following non-linear Poisson equation:
 
\begin{equation}
    \nabla \cdot \bigg(\mu \bigg( \frac{|\nabla \Phi|}{a_o} \bigg) 
  \nabla \Phi \bigg)=4G\pi\rho = \nabla \Phi_{N}
	\label{eq:possion_TeVeS}
\end{equation}

The weak-field metric in TeVeS is the similar to that of GR 
 but when the Newtonian potential 
 is replaced by a total potential $\Phi = \Phi_N + \phi$, 
 where $\phi$ is a scalar field. The equation~(\ref{eq:metric_TeVeS}) 
 also contains a parametrised interpolating function that applies  
 in TeVeS theory for weak and intermediate gravity; 
 for the usual case of $\alpha = 0$ in Eq.46 of \citet{Famaey 2012} 
 this is given by 

\begin{equation}
    \mu(x)= \frac{2x}{1 + 2x + \sqrt{1 + 4x}}
	\label{eq:mu_TeVeS}
\end{equation}
\\
where $x = { |\nabla \Phi| } / {a_o}$, hence 
 giving the same dynamics as MOND with the above $\mu(x)$ 
 in the weak-field limit. 

Also, we note that the recent near-simultaneous detection of
 gravitational waves and the short
 gamma ray burst (GW 20170817 and GRB 20170817A) 
  appears to strongly exclude the standard version of TeVeS 
 \citep{Boran 2017}. 
  However, some versions of modified gravity theories do survive
 this constraint, 
 including the various classes of $f(R)$ and $f(R,T)$ theories 
 {\newtwo (e.g. \citealt{mendoza13}, \citealt{cdl11}), } 
 so other tests as studied below remain 
  potentially valuable.  
\\
\subsection{The External Field Effect (EFE)}
 \citet{Famaey 2012} provide a comprehensive 
 review of alternative theories for the mass discrepancies 
 within the universe, where the observed motions of various  
  galaxy systems 
  exceed the values explained by the mass in visible stars and gas. 
 In practice, no objects are 
 truly isolated in the universe and 
 this has wider and more subtle implications in MOND-like theories 
  than in Newton/GR gravity. 

 Section 6.3 of \citet{Famaey 2012} focuses on the relations 
 between the internal subsystem dynamics and the external parent system 
 gravitational field, commonly known as the external field effect 
 (hereafter EFE).  {\newtwo See also \citet{lfk14} for additional 
 discussion of the EFE; but see results of \citet{Hernandez 2017} and
  \citet{durazo17} for observational hints against the EFE.}  

{\newtwo On the assumption that it is applicable, }  
  the EFE can partly hide most possible MOND-like effects in subsystems 
 such as open clusters within the galactic disk or in wide binaries, 
 apart from a possible rescaling of the gravitational constant.
  However, in the case of main interest here of wide binaries located in the 
 Solar neighbourhood, the galactic EFE (from the baryonic mass 
 distribution in our Galaxy) is quite close to 
 $ g_e \approx 1.0 \,a_0$, so it turns 
 out that the MOND-like effects are considerably 
  reduced but are not fully eliminated by the inclusion of the EFE.  
 
 We note here that the Galactic acceleration for a 
  circular orbit with observed values 
  $v_{LSR} \simeq 220 \kms$ and $R_0 \simeq 8 \kpc$ 
  is somewhat larger than $a_0$,   with $g_{circ} \simeq 1.6 \, a_0$.  
  However, for current estimates of the Galactic stellar mass 
  $\approx 5 \times 10^{10} \msun$ for the disk and $1 \times 10^{10} \msun$
  for the bulge (\citealt{Licquia 2016}; \citealt{McMillan 2017}),   
  the Newtonian contribution from the observed baryonic matter 
   in the Galactic disc and bulge is 
  quite close to $g_{bar} \simeq 1.0 \, a_0$, with the difference generally 
    attributed to DM;  
 in modified-gravity theories 
   without DM, the ratio of these needs to be accounted for by the 
  appropriate modification of gravity via 
   the selected $\mu$ or $\nu$ interpolating function;  
 therefore, it is the smaller value $g_{Ne} = g_{bar}$ which is applicable 
  for the EFE estimates below.  This distinction is notable, since 
  we find below that the fractional difference between MOND-like and 
  Newtonian predictions decreases quite steeply for $g_{Ne} > 1 \, a_0$.  

\subsubsection{Newton/GR dynamics}
 In standard GR,  
  the internal dynamics of an isolated subsystem are independent from 
 the (uniform) external field of the parent system in which it resides, 
  e.g. the internal dynamics of a star cluster  within a galaxy
  are independent of the external uniform gravitational 
 field of the galaxy,  keeping the star cluster in free-fall within the 
  galaxy's frame of reference.
 This is built in as the fundamental Strong Equivalence Principle of GR.
  If the external field varies across the 
 subsystem, this manifests itself as tidal effects, which are 
  rather small in the case here for binaries with $r < 20 \kau$.  

\subsubsection{MOND/TeVeS dynamics}
 Since MOND is an acceleration-based theory,  
  it has to break the Strong Equivalence Principle. 
 What counts is the total gravitational acceleration, 
 with respect to a pre-defined frame (e.g.,the CMB frame).
\footnote{Different MOND theories offer very different answers 
 to the generic question 'acceleration with respect to what?'. 
 For instance, in the MOND-from-vacuum, 
 the total acceleration is measured with respect to the 
 quantum vacuum, which is well defined.} 
 Full MOND effects are thus only observed in systems 
 where the absolute values of the gravitational acceleration, 
 both internal $g_i$ and external $g_{e}$ (e.g. host galaxy, 
 galaxy cluster, etc) are both significantly smaller than $a_0$. 

\subsubsection{EFE dynamics}
The EFE is a remarkable property of various MOND theories, and because this
breaks the strong equivalence principle, it allows us to derive properties 
 of the gravitational field in which a system is embedded 
  from its internal dynamics (and not only from tides). 
 The approximate limiting cases are 
\begin{itemize}
\item $g_i < a_0 \ll g_e$ \text{ - } Newtonian 
\item $g_e < g_i \ll a_0$ \text{ - } Standard MOND
\item $g_i < g_e \la a_0$ \text{ - } quasi-Newtonian with 
  re-normalised gravitational constant, $\Geff$
\end{itemize} 

In the case of interest here, both the internal binary acceleration 
   $g_i$ and the Galactic acceleration $g_e$ are each comparable to $a_0$; 
 this means that there is no simple analytical limit 
 but the acceleration law needs to be estimated 
  numerically, and we see below that the results turn
 out to be somewhat sensitive to the specific version of 
   modified gravity considered.  

Milgrom's gravitational acceleration law, including 
 the EFE is given by:
\begin{equation}
g_{N} = g_i\mu\bigg(\frac{g_i+g_e}{a_0}\bigg) + 
 g_e\bigg[\mu\bigg(\frac{g_i+g_e}{a_0}\bigg) - 
 \mu\bigg(\frac{g_e}{a_0}\bigg)\bigg] 
\label{eq:efe1}
\end{equation}  
which implies that as $g_i \rightarrow 0$ we have Newtonian gravity, 
 $g_{N}$ but with a re-normalised effective gravitational constant,\\ 
\begin{equation} 
\label{eq:gnorm} 
  \Geff \approx \frac{G}{\mu(x)(1 + \frac{d\ln\mu}{d\ln x})}, \ 
   x = g_e / a_0  
\end{equation} 
 Alternatively equation ~(\ref{eq:efe1}) can be re-expressed 
 for the internal gravitational acceleration of the system 
 in terms of Newtonian $g_{N}$, also including the external field:
\begin{equation}
 g_i = g_{N}\nu\bigg( \frac{g_{N} + g_{{Ne}} }{a_0} \bigg) + 
 g_{{Ne}}\bigg[\nu\bigg( \frac{g_{N} + g_{{Ne}} }{a_0} \bigg) 
   - \nu\bigg( \frac{ g_{{Ne}} }{a_0} \bigg)\bigg] 
\label{eq:efe2}
\end{equation}
 where $g_N$ and $g_{Ne}$ are the internal and external Newtonian
 accelerations, $g_i$ is the resulting MONDian internal acceleration, and 
   $\nu(y)$ is the interpolating function
 expressed in terms of the parameter 
  $y \equiv (g_{N} + g_{Ne})/a_0$ or  $y \equiv g_{Ne}/a_0$ respectively. 
\\

 The net result of including the EFE via Eq.~\ref{eq:efe2} 
  is that when $g_i < g_{Ne}$ the 
 modified-gravity effects become similar to a rescaling 
 of the gravitational constant, 
   $g_i = \kappa g_{N}$ with a slowly-varying $\kappa$ which typically
 deviates by of order 10 -- 25 percent from 1, depending on the 
 choice of interpolating function $\mu(x)$ or $\nu(y)$ 
  and the value of $g_{Ne}/a_0$.  
 Thus the MOND effects are no longer large, but are still appreciable. 

 
\subsection{Emergent Gravity}
The Emergent Gravity theory originates from the concept of treating 
 gravity as an entropic force. The theory of Emergent Gravity has 
 recently been developed by \citet{Verlinde 2016} and  
 \citet{Hossenfelder 2017}. Emergent Gravity is 
 the notion of describing the macroscopic 
   nature of spacetime (aka GR) ``emerging'' from an underlying 
  microscopic description of spacetime. (See however \citep{dai-stoj} for
 some potential problems with this formulation). 
\\
The emergent nature of spacetime is postulated to stem from the 
 thermodynamic laws for a black hole, centered
  around the Bekenstein-Hawking entropy \& Hawking temperature, expressed as;
\begin{equation}
S_{BH} = \frac{A(r)}{4G\hbar} \\
T_{BH} = \frac{\hbar \tilde{a}}{2 \pi}
\end{equation}
where $A(r)$ is the area of the horizon and $\tilde{a}$ is the 
 surface acceleration. 

The Bekenstein-Hawking entropy can determine the amount of 
 quantum entanglement (QE) in a vacuum, where QE plays a role in 
 explaining the connectivity of classical spacetime. 
 From this notion, one can use the theoretical concept of 
 linking quantum information theory (QIT) with the emergent spacetime.
The theoretical framework is in representing the spacetime  
 geometry as a QE structure, governed by an entropic description 
 (hereafter, entropic QE).
The spacetime vacuum is made-up of a network 
 of QE units of quantum information (QI), 
 which are the fundamental microscopic constituents of spacetime 
 (i.e., QE units bond together creating a network; 
 and this network of QE units is what spacetime is made up of). 
 Matter \& energy are proposed to influence the microscopic 
 constituents (or the QE unit structure embedded 
 within spacetime itself), resulting in 
 a macroscopic curvature of spacetime.

\subsubsection{Apparent DM in emergent gravity}
Emergent Gravity describes dark energy (DE) and
  the apparent dark matter (DM) to have a common origin both 
  connected to the emergent nature of spacetime. 
 It also results that the flattening of galaxy rotation curves 
  are controlled by the Hubble acceleration scale,  
\begin{equation*}
 a_\Lambda \approx cH_0 = 	\frac{c^2}{L}
\end{equation*}
where $c$ is the speed of light, $H_0$ is the Hubble constant, 
 $L$ the Hubble length and $a_\Lambda$ is an acceleration constant.
\footnote{Note that Verlinde (2016) uses symbol $a_0$ for this, but it is
  different (by roughly a factor 5) from the usual MOND parameter
 $a_0$ used above; so we have used symbol $a_\Lambda$ replacing
   Verlinde's $a_0$.}   
Since $H$ is actually time-dependent we should replace this
 with $a_\Lambda = c H_0 \sqrt{\Omega_\Lambda}$ to get a time-invariant
 parameter; this has the 
 appealing feature that the acceleration scale $a_\Lambda$
 is naturally related to the observed value of the cosmological
 constant (unlike standard MOND where the constant $a_0$ is 
  an arbitrary free parameter).  

The apparent effects of DM appear 
  at scales below $a_\Lambda$, equivalent to a surface mass density 
\begin{equation}
\Sigma(r) = \frac{M}{A(r)} < \frac{a_\Lambda}{8 \pi G} 
\end{equation} 
In terms of entropy, 
\begin{equation}
S_M = \frac{2 \pi M}{\hbar a_\Lambda} < \frac{A(r)}{4G\hbar} \ . 
\end{equation}
The involvement of DE in Emergent Gravity is associated with 
 the entropic description of the QE structure. 
 This association describes the 'stiff' geometry of spacetime 
  manifesting into an elastic nature of spacetime 
 at scales below $a_\Lambda$ .
  The elastic response of the DE ``medium" takes 
 the form of an extra apparent dark force which then gives rise to
 the effects which are normally attributed to DM. 

\subsubsection{Covariant version of Emergent Gravity}
 In the work of \citet{Hossenfelder 2017}, 
  Emergent Gravity is constructed in a Lagrangian form, 
 showing the underlying mechanisms, 
 within a de-Sitter space filled with a vector-field that couples to 
 baryonic matter and, by dragging on it, creates an effect similar to DM. 
 Also, the vector-field mimics the behaviour of DE treating the 
 spacetime as an elastic medium. The theory of Emergent Gravity
  interprets between the gravitational equations with linear 
 elasticity equations  (i.e., relating gravity quantities with elastic 
 quantities), (see Section 6 in \citep{Verlinde 2016}). 
 From \cite{Hossenfelder 2017} the action for Emergent Gravity is expressed as:
\begin{equation}
S_T = S_{EH} + S_{int} + S_{\theta} + S_{M} 
\label{eq:EMRGNT}
\end{equation}
Where $S_{EH}$ is the Einstein-Hilbert action and $S_{M}$ is the action 
 for matter fields, see equation~(\ref{eq:E-H action}); 
 $S_{int}$ \& $S_{\theta}$ are the self-interaction and the 
 imposter field actions\footnote{Note $S_{int}$ and $S_{\theta}$, the 
 self-interaction and the imposter field actions are yet to have a 
 definitive description due to the theory being recently developed}, 
 expressed as:

\begin{equation}
S_{int} = \int \frac{-u^{\mu} n^{\nu}}{L}T^{\mu \nu} \, d^4x \\ = 
  \int \frac{u^{\mu} u^{\nu}}{L u}T^{\mu \nu} \, d^4x 
\end{equation}
\begin{equation}
S_{\theta} = \int \frac{M_{p}^2}{L^2}\chi^{\frac{3}{2}} - 
 \frac{\lambda^2 M_{p}^2}{L^4}u_ku^k d^4x
\end{equation}
These actions describe the elastic behaviour and force of the spacetime 
 geometry of the theory.  Also 
\begin{equation}
\chi = \frac{-1}{4}\epsilon_{\mu \nu}\epsilon^{\mu \nu}+ \frac{1}{3}\epsilon^2 
\end{equation}
  is the kinetic term for the vector fields, and  
\begin{equation} 
\epsilon_{\mu \nu} = \nabla_{\mu}u_{\nu} + \nabla_{\nu}u_{\mu}  
\end{equation}
is the strain tensor. 

\subsubsection{Newtonian weak-field limit}

In the context of GR, astrophysical systems such as galaxies are dominated 
 by DM, resulting in an approximately flat rotation curve. 
 In the context of Emergent Gravity, these systems are described 
  as the baryonic matter 
 reducing the amount of entropic QE structure of the surrounding spacetime,
  while in the regions where is negligible matter 
 where the acceleration is $a \leq a_\Lambda$, the 
 spacetime manifests into an elastic DE medium, 
 where the elastic response of this medium results in an 
 extra 'dark force', which mimics the effects of DM in the outer 
 regions of galaxies. \\
This force can be computed via entropy:

\begin{equation}
S_M(r) = \frac{-2\pi Mr}{\hbar}
\label{eq:sm}
\end{equation}
$S_M(r)$ is the amount of entropic QE structure 
 that the mass $M$ has removed 
 from a region of size $r$. 
 This can also be expressed in terms of volume:
\begin{equation*}
S_M(r) = \frac{-V_M(r)}{V_0}\ , \quad V_M(r) = \frac{8\pi G Mr}{a_\Lambda 
  (d-1)} \ ,  \quad  V_0 = \frac{4G\hbar L}{(d-1)}
\end{equation*}
 where $V_0$ is the volume per unit entropy, $V_M(r)$ is  
 volume containing the amount of entropy that has been 
  reduced by mass $M$ from a region of size $r$, and $d$ 
 is the number of dimensions within the theory derived from the 
 gravitational \& elastic dynamics correspondence, 
 (See section 6 of \citet{Verlinde 2016}).
\\
For the remaining spacetime region containing no baryonic matter is given by:
\begin{equation}
S_{DE} = \frac{V(r)}{V_0} = \frac{rA(r)}{L 4G\hbar} =
   \frac{r a_\Lambda A(r)}{4G\hbar}
\label{eq:sde}
\end{equation}
 where $S_{DE}$ is the total entropic QE associated with DE,
 treating the spacetime as an elastic medium (or entropy of the DE medium), 
 $V(r)$ is the volume of the whole system,
  which also contains $V_M(r)$.(e.g., a galaxy containing 
 both visible matter and DM). 
\\
Taking the ratio between equation~(\ref{eq:sm}) \& equation~(\ref{eq:sde}), 
 one can obtain the equation for the surface mass density, given by:
\begin{equation}
\Sigma(r) = \frac{a_\Lambda}{8\pi G}\epsilon(r)
\label{eq:smdde}
\end{equation}
where $\epsilon(r)=\frac{S_M}{S_{DE}}=\frac{V_M(r)}{V(r)}$ 
 is the strain tensor representing the transition from Newton/GR 
 to the emergent DE medium elastic effect.\\
If $\epsilon(r) > 1$, all entropic QE structure is 
 reduced by matter, leading to the usual Newtonian/GR dynamics. 
 For $\epsilon(r)\leq 1$ the regions without matter, 
 the spacetime results in a DE elastic effect 
 (or extra dark force) modifying Newtonian/GR dynamics. 
\\
\\
Since we are dealing with very low acceleration 
 regimes, in the case of $\epsilon(r)\leq 1$, we can obtain the 
 gravitational acceleration of Emergent Gravity on the extremely weak scale.
\begin{eqnarray}
\big(\frac{8\pi G}{a_\Lambda}\Sigma(r)\big)^2 &=& \epsilon(r)^2 \\
\epsilon(r)^2  & = & \frac{1}{A(r)}\frac{dV_M(r)}{dr}  \nonumber \\ 
  &= & 
  \frac{8\pi  GM}{a_\Lambda A(r)(d-1)}=\frac{8\pi G}{a_\Lambda(d-1)}\Sigma_M(r)
\end{eqnarray}
where $\Sigma_M(r) = \Sigma_B(r)$ is the baryonic surface mass density in the  
 region where baryons reduce the entropic QE structure.\\ 
From Section 6 of \citet{Verlinde 2016}) we can use the
  surface-mass-density and gravitational interaction relation:
\begin{equation}
\Sigma_i = \frac{(d-2)}{(d-3)} \frac{g_i}{8\pi G}
\label{eq:smdrel}
\end{equation}

Taking the RHS of equation~(\ref{eq:smdde}) to be $\Sigma_{DE}(r)$, 
 the remaining elastic DE medium surface mass density 
 where the "DE medium" elastic response (or dark force) takes effect, 
 applying equation~(\ref{eq:smdrel}), 
 one can obtain the relation between $\Sigma_{DE}(r)$ \& $\Sigma_{M}(r)$ 
 expressed as:
\begin{equation}
\Sigma_{DE}(r)^2 = \frac{a_\Lambda}{8\pi G(d-1)}\Sigma_{M}(r)
\label{eq:smdrel1}
\end{equation}
and the gravitational acceleration relation of the elastic response of 
 the DE medium (or dark force) can be expressed as 
\begin{equation}
g_{DE} = \sqrt{a_m g_{B}} \\ a_m = \frac{a_\Lambda (d-3)}{(d-2)(d-1)}
\end{equation}
\\
 and thus in the case $d = 4$ gives $a_m = a_\Lambda/6$, in 
  which case $a_m$ is numerically 
  rather close to the usual MOND value 
 $a_0 \simeq 1.2 \times 10^{-10} \msecsq$ above. 

The total gravitational acceleration of a whole system in Emergent Gravity 
 is given by: 
\begin{equation}
 g_{{EG}} = g_{B} + g_{{DE}} = g_{N} + \sqrt{a_m g_{N}}
\end{equation}
where $g_{B} = g_{N}$ is the Newtonian acceleration. 

\section{Simulations of Orbits} 
\label{sec:orbs} 

Previously in Section~\ref{sec:wb-gaia} we estimated 
 statistical distributions of $u_{3D}$ 
 in the idealised case of Newtonian gravity where both masses
 and all six relative position and velocity
 components are known (but accelerations are not measurable).  
 We next use numerical orbit simulations to deal with the 
  cases of orbits in modified-gravity theories, and also the more
 observationally realistic case where the 
  projected separation of the binary is well measured 
 but the radial component of separation is unknown (or only 
  has an upper bound), as below. 

\subsection{Orbit simulations with modified gravity} 
\label{sec:orbs2} 
 Here we simulate a large sample of $\sim 5 \times 10^6$ orbits
 with random values of $a, e$ in each of the various gravity theories 
  outlined above, then study the joint distribution of observables, 
 in particular projected separation 
  $r_p$ and relative velocity $\vthree / v_c(r_p)$, where 
  $v_c(r_p)$ is the Newtonian velocity for a circular orbit at
  the {\em current} projected separation.  
  The latter is readily 
 calculable given the estimated masses of both binary components, 
  and the resulting dimensionless ratio is convenient since
  the distribution should be independent of $r_p$ in the case
  of Newtonian gravity when the eccentricity distribution 
   $f(e)$ is independent of $a$, and should have 80th/90th 
  percentile values nearly independent of $f(e)$.  

 In the case of modified gravity, the orbits are generally 
 not closed ellipses, so they are not strictly defined by 
 the standard Keplerian parameters $a, \, e$, but we still need
  to simulate a distribution in size and shape of orbits.  
  To deal with this, for a modified-gravity orbit 
   we define an ``effective'' orbit size $\hat{a}$ and quasi-eccentricity 
   $\hat{e}$ as follows: we define 
   $\hat{a}$ to be the separation at which the simulated relative velocity
   is equal to the circular-orbit velocity (in the current 
   modified-gravity model), then
   we define $\theta_{\rm circ}$ to be the angle between the relative velocity 
   vector and the tangential direction when the orbital separation crosses
    $\hat{a}$, 
  and then $\hat{e} \equiv \sin \theta_{\rm circ}$;  these definitions 
  coincide with the usual $a, e$ in the case of standard gravity. 

We simulate orbits for both Newton and various MOND cases,  
  using a fourth-order Runge-Kutta integration. 
 For the orbit initial conditions we take initial separation 
 $r(t_0) = \hat{a}$, total relative velocity $v_0 = \sqrt{g_{Grav} \, r_0}$
  where $g_{Grav}$ is the relative acceleration in the given model, 
 and angular velocity $\dot{\theta}_0 = (v_0/r_0) \cos \hat{e}$ and
 radial velocity $\dot{r}_0 = v_0 \sin \hat{e}$;  
  where $0\leq \hat{e} \leq 1$ is the (pseudo-)eccentricity of the orbit. 
 We adopt a simulated 
 eccentricity distribution given by $f(\hat{e}) = 0.4 + 1.2 \, \hat{e}$, 
  as estimated for wide binaries by \citet{Tokovinin 2016}. 
\\
After integrating these orbits using one of a selected set
 of gravity laws (Newton/GR, MoND, etc) and a chosen
 value for external field $g_e$,  we ``observe'' the
 resulting binaries at many random times and random inclinations to the 
  line-of-sight.  
 For each simulated orbit/epoch snapshot, we produce simulated 
 observables including the projected separation $r_p$, 3D relative
 velocity $\vthree$, and also $\vthree/v_C(r_p)$ as the ratio to 
  the circular velocity (for Newtonian gravity) 
 at the instantaneous projected separation,  
  where $r_{p} = r \sin\beta$ is the projected separation 
 of the orbit and $\beta$ is the angle between the binary separation  
  and the line of sight. 
\\

The radial acceleration law is chosen according to the
 selected gravity theory under consideration, and also with the
 external field effect turned off or on (see below).  
 For the Newtonian/GR case, we have the standard 
\begin{equation}
g_{N} = \frac{G(M_1+M_2)}{r^2}
\label{eq:g_n}
\end{equation}
\\
For the MOND case with the ``simple'' interpolating function, we have 
\begin{equation}
g_{{M1}} = \frac{g_{{N}}}{\mu(\frac{g_{{M1}}}{a_o})} \ , \\ 
 \mu(x) = \left( 1+ \frac{1}{x} \right)^{-1} \ ; 
\label{eq:g_mond1}
\end{equation}
 this interpolating function 
 is actually known to predict excessive deviations in the Solar system 
  \citep{Famaey 2012}, but does provide 
  a good fit to galaxy rotation curves in the $x \la 1$ regime, and it could 
  readily be modified to converge faster to $\mu(x) \rightarrow 1$ 
  at large $x \gg 1$ to avoid the conflict with Solar system observations.   
\\
For the MOND case with the ``standard'' interpolating function, we have 
\begin{equation}
g_{{M2}} = \frac{g_{{N}}}{\mu(\frac{g_{{M2}}}{a_o})} 
  \\ \mu(x) = \frac{x}{\sqrt{1 + x^2} } \ ; 
 \label{eq:g_mond2}
\end{equation}
 this function converges faster to $1$ at large $x$, 
  though it provides a somewhat less good fit to galaxy rotation curves 
   since the transition from modified to Newtonian 
 gravity around $x \sim 1$ is rather abrupt \citep{Famaey 2005}.   
\\
We also use the fitting function of \citet{McGaugh 2016} (hereafter MLS), 
  sometimes known as the ``mass discrepancy acceleration relation'', 
 given by 
\begin{equation} 
 g_{MLS} = g_N \nu(g_N/a_0) \ ; \\ \nu(y) = \frac{1}{1 - \exp(-\sqrt{y})} \ ; 
\label{eq:g_mls} 
\end{equation} 
 we refer to this as the MLS interpolating function below. 
 This function is shown by MLS to produce a good fit to rotation curves
  for a large sample of disc galaxies spanning a range of masses;
 it also has the feature that the function $\nu(y)$ converges
  very rapidly to 1 when $y \ga 20$, so deviations on Solar System 
  scales are predicted to be vanishingly small.  

For the TeVeS case we adopt 
\begin{equation}
g_{T} = \frac{g_{N}}{\mu( {g_{T}}/ {a_o}) } 
 \\ \mu(x)= \frac{2x}{1 + 2x + \sqrt{x^2 + 4x}} 
	\label{eq:g_teves}
\end{equation}
\\
For the Emergent Gravity case we adopt 
\begin{equation}
g_{EG} = g_{N} + \sqrt{a_m g_{N}}
 \label{eq:g_eg}
\end{equation}

To apply the External Field effect (EFE), we use 
\begin{equation}
g_{i, EFE} = g_{N}\nu\bigg( \frac{g_{N} + g_{Ne} }{a_0} \bigg) + 
  g_{{Ne}}\left[\nu\bigg( \frac{g_{N} + g_{Ne} }{a_0} \bigg)  
   - \nu\bigg( \frac{ g_{Ne} }{a_0} \bigg) \right]
\label{eq:g_efe} 
\end{equation}
 where $g_N$ is internal Newtonian acceleration, $g_{Ne}$ is the external 
 (Galactic) Newtonian acceleration, and $g_{i, EFE}$ is the ``true''
  internal acceleration with application of the external field effect. 
 We ran simulations with three selected values of the external-field
 acceleration $g_{Ne}$, respectively
\begin{equation*}
 g_{Ne} = \text{[0.5, 1, 1.5]} \, a_0 
\end{equation*}
  which bracket the values for the local Galactic acceleration.   

We simulate orbits using each of the above $g$ formulae in 
  Eqs.(\ref{eq:g_n} -- \ref{eq:g_eg}), 
 with a flat distribution in $\ln \hat{a}$ and a Tokovinin
  distribution for $\hat{e}$, and integrate the orbits in time 
 using the RK4 integration. 
   These simulated binaries are then ``observed'' at random phases
 and inclination angles, with the results discussed below. 

\subsection{Results of simulated orbits}
\label{sec:results} 
 
 We show results of simulated observables for the orbits in various 
  cases of gravity model in Figures~\ref{fig:hist35-noefe} to 
  \ref{fig:hist710-efe-mls} below.  
 In each case the main observable parameters of interest 
  are the projected separation
 $r_p$ and the velocity ratio $\vthree /v_C(r_p)$, i.e. the ratio  of
 the 3D relative velocity to that of a circular (Newtonian) orbit
  at the current projected separation.   We show 
  histograms of $\vthree/v_C(r_p)$ in selected bins of projected
  separation, normally with bin widths of a factor of $\sqrt{2}$ 
  in projected separation.   

\begin{figure*}
\hspace*{-1mm}\includegraphics[width=15cm]{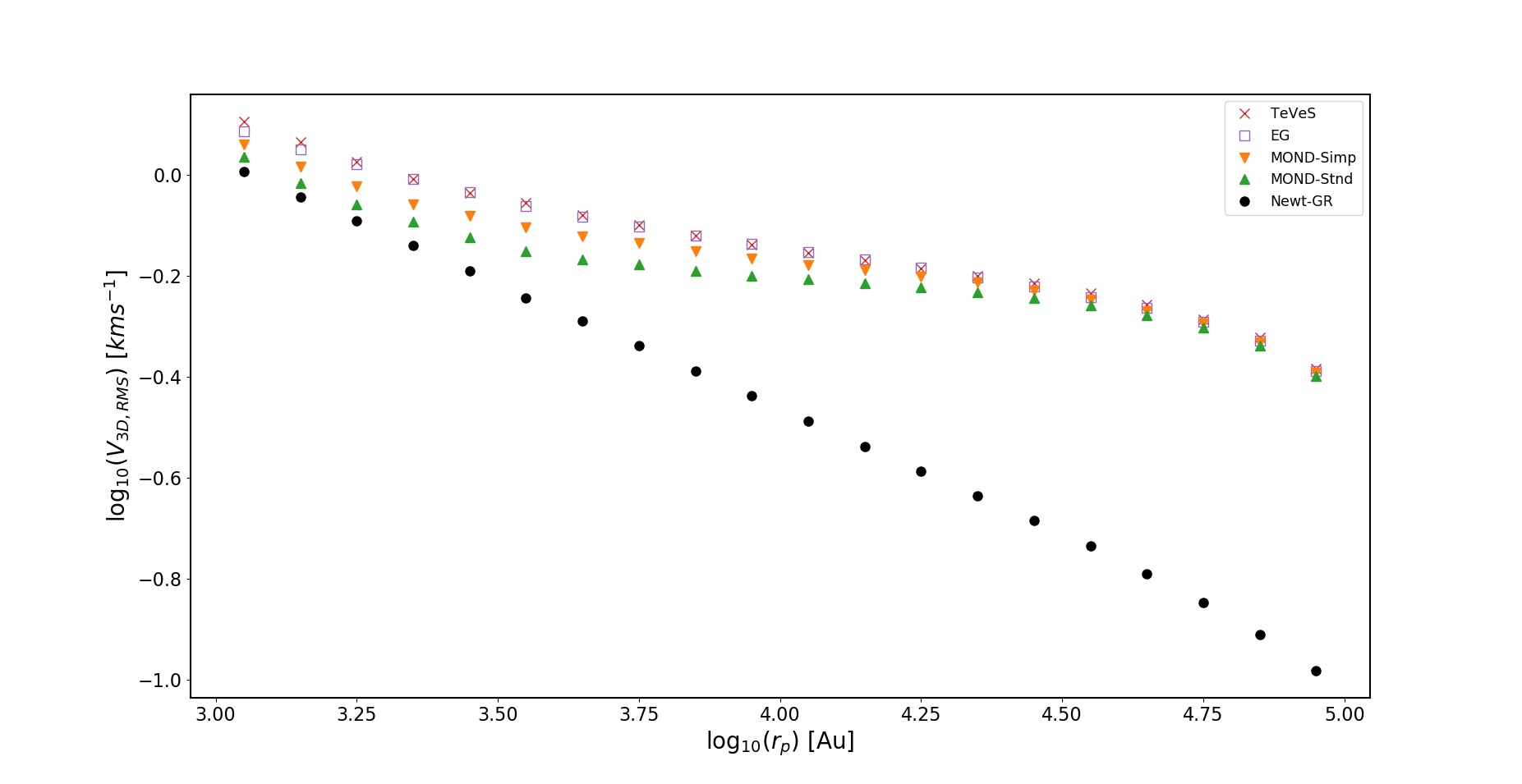} 
\vspace*{-1mm} 
\caption{ Points show the RMS 3D relative velocity
  for the set of simulated binary orbits binned in projected separation $r_p$; 
   for orbits {\em excluding} the external field
  effect in MOND-like theories. As in the legend, from bottom to top: 
   black circles show Newtonian gravity;
   upward (downward) pointing triangles show MOND standard (simple) 
  interpolating function respectively; open squares show emergent gravity; 
  crosses show TeVeS. 
  \label{fig:vrms-noefe} }
\vspace{2mm} 
\end{figure*}

\begin{figure*} 
\hspace*{-1mm}\includegraphics[width=15cm]{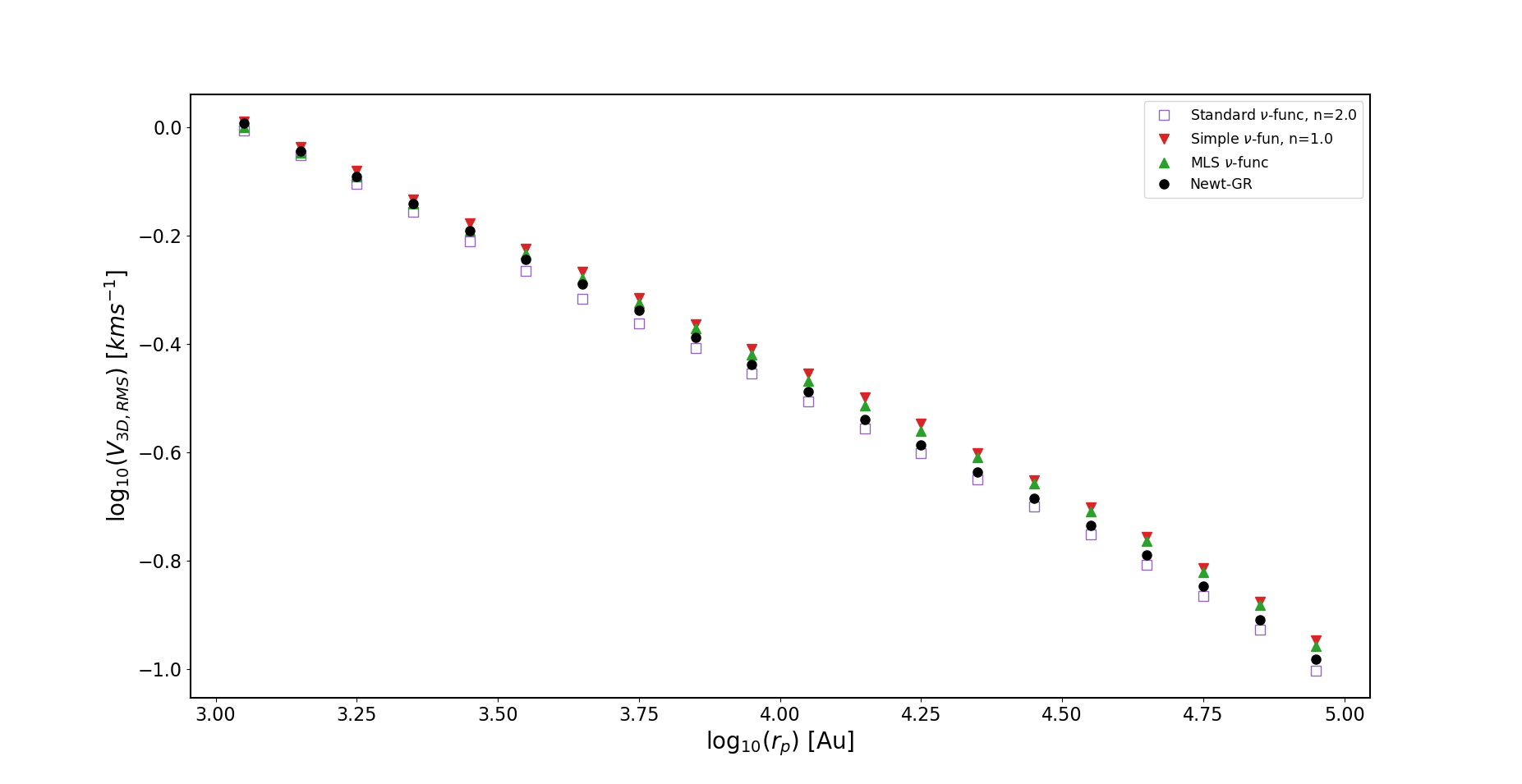} 
\vspace*{-1mm} 
\caption{ Points show the RMS 3D relative velocity
  for the set of simulated binary orbits  binned in projected separation $r_p$; 
  for orbits {\em including} the external field
  effect all with $g_{ext} = 1.0 \, a_0$. 
    As in the legend, 
  black points show Newtonian gravity; 
   open squares show MOND standard interpolating function; 
 downward-pointing triangles show MOND simple interpolating function; 
  upward-pointing triangles show MLS interpolating function. 
  \label{fig:vrms-withefe} }
\end{figure*}

{\newtwo Partly for comparison with previous work 
  (e.g. \citealt{jiang-tremaine}, \citealt{Hernandez 2012}), 
  we show the root-mean-square 3D velocity difference evaluated 
  in bins of  projected separation for various gravity models.  
  This is clearly a simple  statistic, but is not necessarily 
  optimal for real-world application  since it is well known that
  RMS statistics are rather non-robust to outlier contamination e.g. from
   unbound or fly-by pairs.   
 Figure~\ref{fig:vrms-noefe} shows the MOND-like models without the EFE,
  producing rather substantial deviations above Newtonian.
  ( The Newtonian case
  shows the expected decline as $RMS(v_{3D}) \propto r_p^{-0.5}$, 
  except for a small turn-down below this 
  at projected separations $\ga 50 \kau$; the latter 
 is due to our truncation of orbits with apocentre 
  beyond $300 \kau$). 
  Results with the EFE included (for $g_e = 1.0 \, a_0$) are shown in  
  Figure~\ref{fig:vrms-withefe}. This shows that with the EFE included
  the deviations are much more subtle, and essentially
  saturate at a near-constant multiplicative offset from Newtonian
   at separations $r_p \ga 10 \kau$. The size of the offset 
   is less than 10 percent, and is rather sensitive to
   the choice of interpolating function;  
   see discussion below for potential statistical tests. 
}  




Turning to the histograms of $v_{3D}/v_c(r_p)$, 
 we find that Newtonian gravity predicts that
 the histogram of $v_{3D}/v_c(r_p)$ for wide binaries
  should exhibit a steep decline above values $\sim 1.1$, 
  with an 80th percentile near 1.02 and a 90th
 percentile near 1.14; these features have 
  rather weak dependence on the poorly known 
  distribution of orbit eccentricities.  

 In Figures~\ref{fig:hist35-noefe} to \ref{fig:hist710-noefe}
   we show histograms of $\vthree / v_C(r_p)$ in several selected 
  bins of projected
 separation,  comparing Newtonian/GR gravity and various modified-gravity
 theories {\em without} the EFE.  It is clear from these  
 that all modified-gravity theories {\em without} the EFE produce a large 
 and obvious shift in the distribution,
 with the 90th percentile reaching $\sim 2.0$
   in the
  projected separation bin $(5, 7.1) \kau$.  The specific size
  of the shift is slightly dependent on the modified-gravity
 model considered, but all the modified-gravity models without EFE
  show large shifts: 
 such large effects should be readily detectable by observations, 
   and not reasonably produced by any combination of observational error or 
  sample contamination.  
 We thus conclude that essentially 
  all modified-gravity theories {\em without} an EFE can
 be robustly tested or ruled out by \GAIA wide-binary samples
  with ground-based radial velocity followup. 

\begin{figure*} 
 \includegraphics[width=13cm]{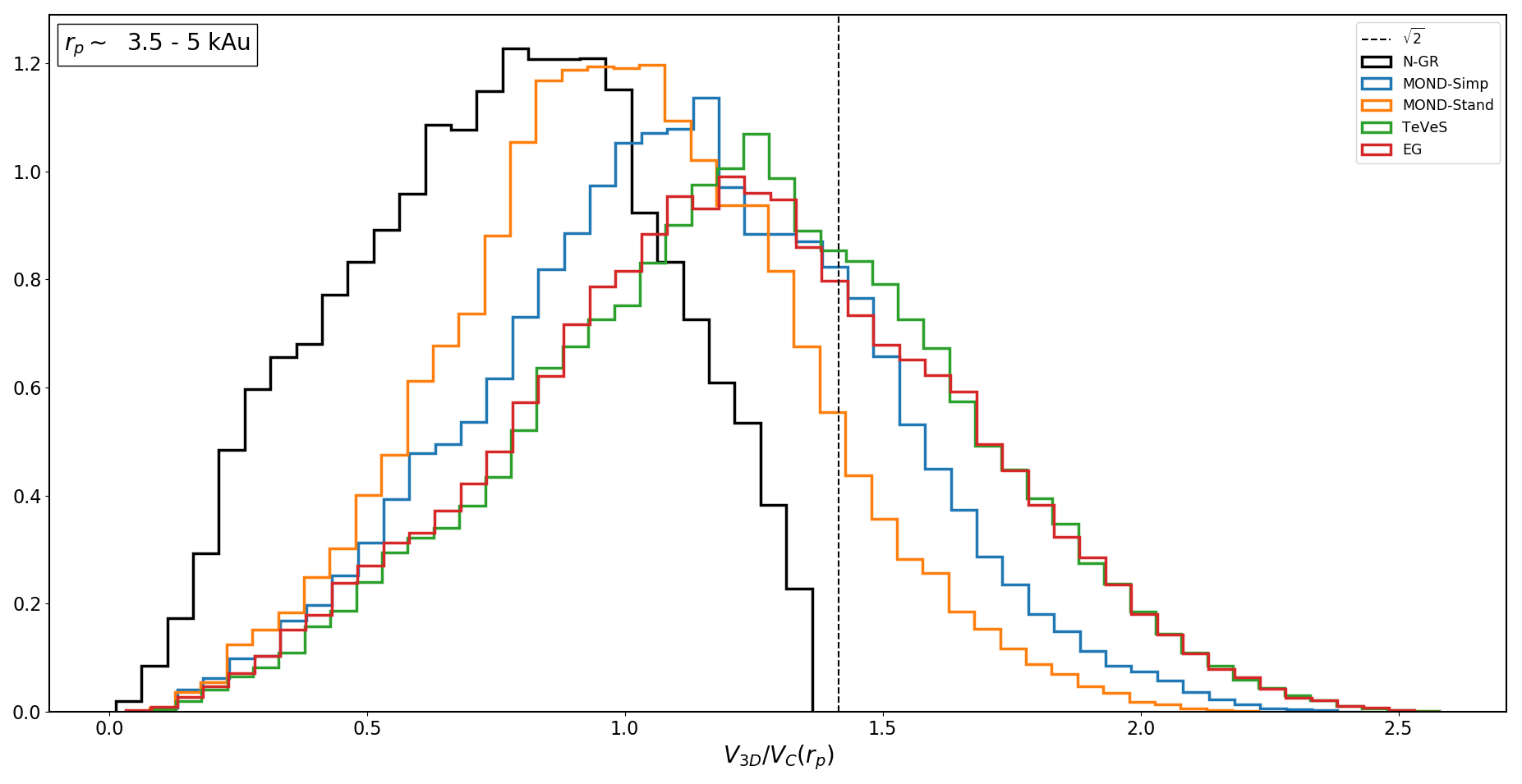} 
\caption{Histogram of simulated $\vthree/v_C(r_p)$ for
 projected separation range $r_p \in (3.5, 5) \kau$, 
 for various gravity models without EFE: Newtonian (black), 
  MOND-simple (blue), MOND-standard (orange), 
    TeVes (green), Emergent gravity (red). 
 \label{fig:hist35-noefe} 
}
 \includegraphics[width=13cm]{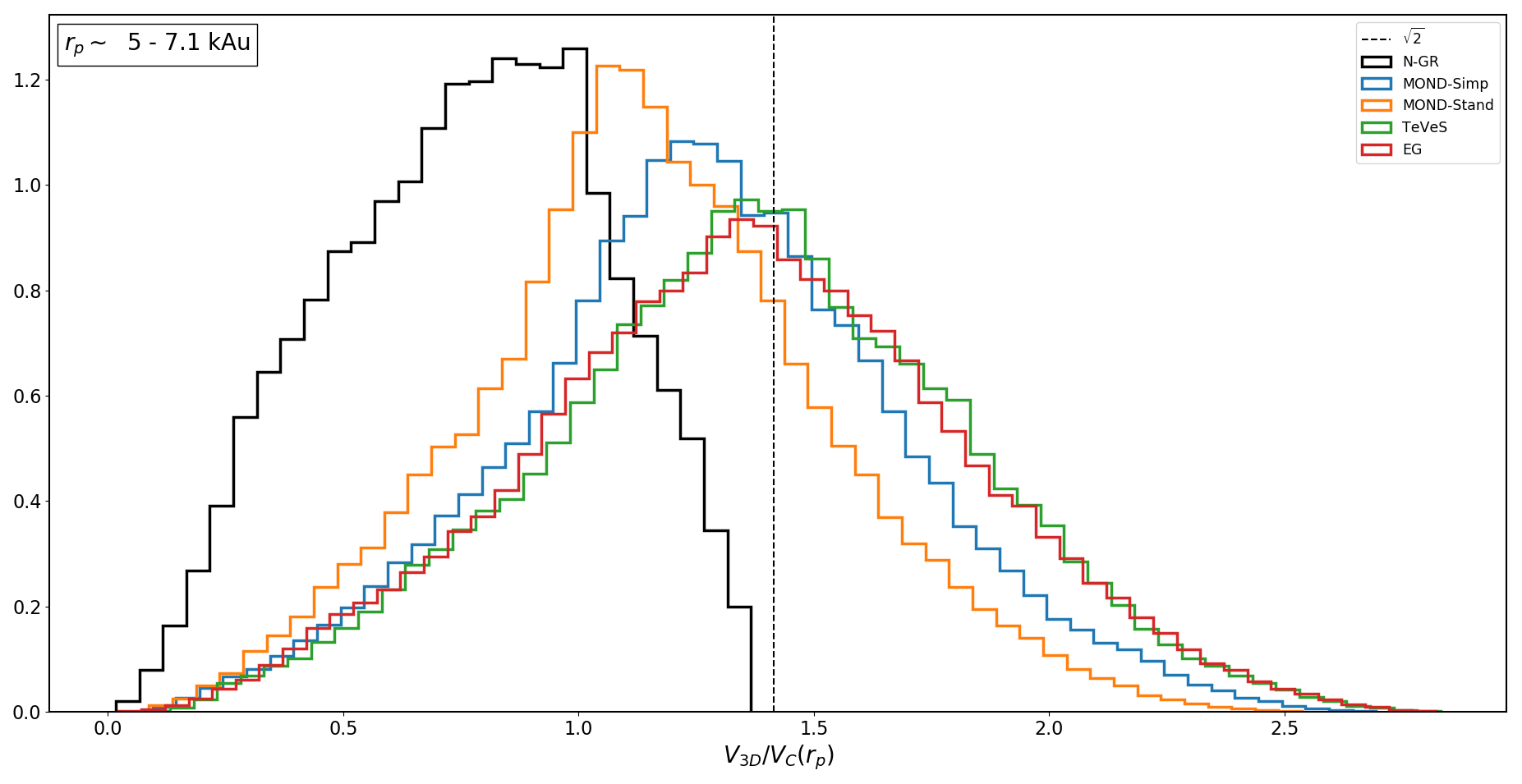} 
\caption{As Figure~\ref{fig:hist35-noefe} for projected
 separation range $r_p \in (5, 7.1) \kau$. 
 \label{fig:hist57-noefe} 
} 
 \includegraphics[width=13cm]{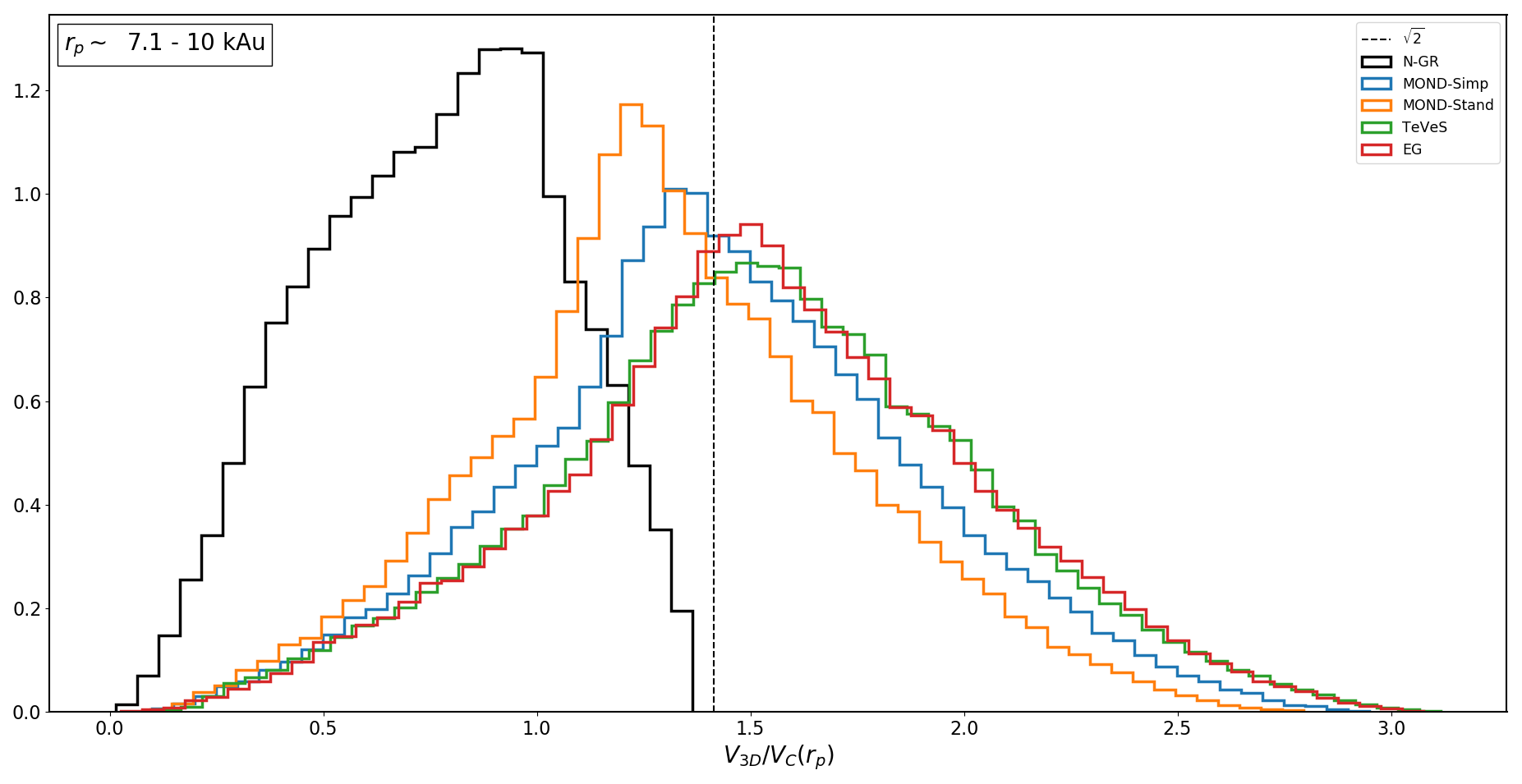} 
\caption{As Figure~\ref{fig:hist35-noefe} for projected
 separation range $r_p \in (7.1, 10) \kau$. 
 \label{fig:hist710-noefe} 
} 
\end{figure*}  

 The next set of Figures show some modified-gravity models 
 with the EFE included: Figures~\ref{fig:hist35-efe-simp} to 
  \ref{fig:hist710-efe-simp} show results for  
  MOND with the Simple interpolation function for different values of external
 field, $g_e = (0.0, 0.5, 1.0, 1.5) a_0$ respectively; so the $g_e = 0$
  case reproduces that from the preceding Figures.  

 Likewise, Figures~\ref{fig:hist35-efe-std} to \ref{fig:hist710-efe-std} 
  show the same bins and $g_e$ values for 
  the case of the MOND standard interpolating function, 
 while Figures~\ref{fig:hist35-efe-mls} to \ref{fig:hist710-efe-mls} 
  show the same bins and $g_e$ values for the MLS interpolating
 function of Eq.~\ref{eq:g_mls}.  
 
 It is seen from these figures that with 
  the EFE included, the shifts in the distributions (relative to Newtonian)  
   are considerably smaller than the no-EFE cases, but 
 the shifts remain non-negligible. The shifts are only marginally
 visible on the histograms, but the shifts particularly in the
 upper percentiles of the distribution remain non-negligible. 
    For the case $g_{Ne} = 1.0 \, a_0$, we show the resulting 
 80th and 90th percentile values in Figure~\ref{fig:perc80}.  These
  reveal that there are offsets relative to Newtonian gravity 
   by approximately 0.04 to 0.08 in these
  percentiles, depending on the MOND function, with the offset 
  increasing slowly with separation.   
 
In essence these offsets occur
 because the EFE leads to approximately a moderate re-normalisation 
 of the effective $G$ at low accelerations:
 i.e. with the EFE included and external acceleration $g_{Ne} \sim 1 \, a_0$, 
   the ratio $g_{i} / g_{Ni}$ is slowly varying with scale
   but is different from 1 at accelerations $\la a_0$, 
 so the EFE leads to approximately quasi-Newtonian gravity but with a re-scaled
  apparent value $\Geff$ of the gravitational constant.     
  The limiting ratio $\Geff /G_N$ is dependent on the choice of MOND 
  interpolating function and the selected value of $g_e$.  
  
  For the ``simple'' MOND interpolation 
 function Eq.~\ref{eq:g_mond1} and $g_e = 1 \, a_0$,  
  we find an upward shift of about 7 percent in the 80th and 90th percentile
  values. 
 
 For the MLS interpolation function Eq.~\ref{eq:g_mls},
  the shift is slightly smaller than the MOND-simple interpolation
 function, with about 4 percent
  shift; this is in the direction expected since the MLS
 interpolation function with external $g_e = 1 \, a_0$ predicts 
   $g_i / g_N \sim 1.09$ for $g_i \la 1 a_0$.  
 
 There is a somewhat unexpected result from the EFE using  
  the ``standard'' MOND interpolation function Eq.~\ref{eq:g_mond2};  
  here the shift is smaller, but actually in the {\em opposite} 
 sense i.e. the 80th and 90th percentiles for $\vthree/v_C(r_p)$ 
  are shifted to marginally {\em smaller} values than the Newtonian case.  
 This rather counter-intuitive result is actually caused by an
   odd ``feature'' of 
 the standard interpolating function including the EFE:
   in the regime where $g_e \approx 1 \, a_0$ and $g_i \la 1 \, a_0$, 
  the EFE with the standard interpolating function 
  actually predicts internal accelerations $g_i$ about 7 percent {\em 
   weaker} than Newtonian, $g_i / g_{Ni} \sim 0.93$, 
  and this fractional suppression is rather 
   slowly varying for $g_{Ni}$ between 0 to $1 \, a_0$.   

 Another point of note is that the shift in the distributions 
  generally becomes apparent at projected separations somewhat smaller 
   than simply the scale $\sim 7 \kau$ where
 the circular-orbit acceleration is comparable to the MoND $a_0$ constant. 
 This occurs for a combination of two reasons: 
   partly because MOND-like effects are expected to become non-negligible 
  at internal accelerations somewhat larger than $1\, a_0$  (as preferred   
  to give near-flat galaxy rotation curves and a reasonably smooth
 interpolation function);  
 and also because the tail of binaries with larger values of 
   $\vthree/v_C(r_p) \ga 1$ is dominated by moderately eccentric orbits 
  which happen to be observed near
  pericenter: therefore at a given projected separation $r_p$, 
  the faster binaries are those with time-average separation larger  
  than the present-day value, hence their past orbit has mainly 
   sampled wider separations where
  the MOND-like effects are relatively larger. 

 This feature is interesting, since it implies that MOND-like effects should 
  already start to become measurably large at projected 
  separations $\sim 3 - 5 \kau$, 
  a range where the survival probability for wide binaries 
  is predicted to be high, the perspective-rotation effects
 (Section~\ref{sec:perspec}) are smaller, and the relative velocities
  are not very small; all of these are favourable from an observational 
  perspective.   


\begin{figure*} 
 \includegraphics[width=13cm]{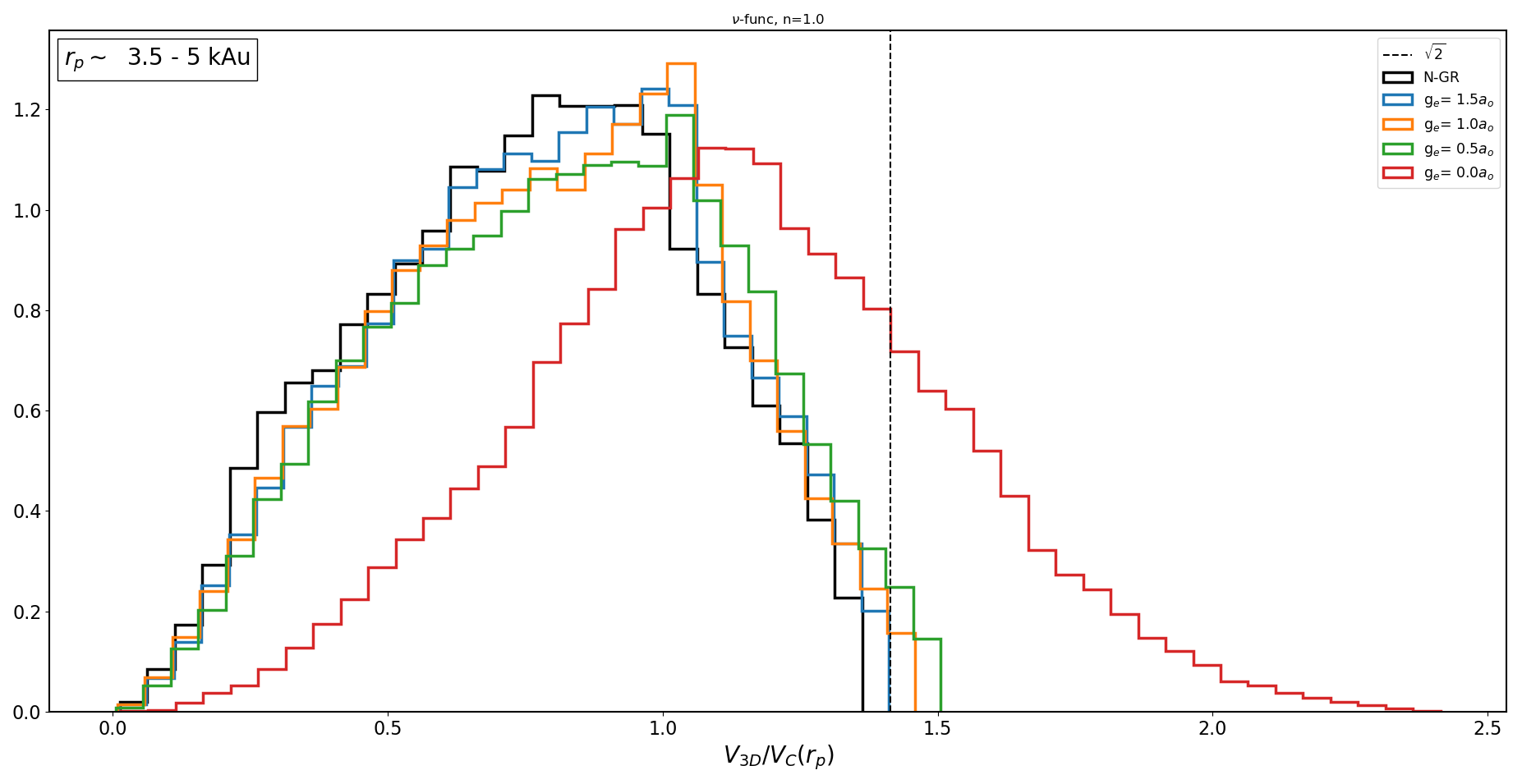} 
\caption{Histogram of simulated $\vthree/v_C(r_p)$ for MOND-simple
  interpolating function, with EFE, for
 projected separation range $r_p \in (3.5, 5) \kau$.  
 Black histogram is for standard Newtonian gravity, and 
 coloured histograms are for MOND simple interpolating function including 
  EFE with four values of external field $g_{Ne}$: $g_{Ne} = 0$ (red), 
  0.5 (green), 1.0 (yellow) and 1.5 (blue) in units of $a_0$.  
 \label{fig:hist35-efe-simp} 
}
 \includegraphics[width=13cm]{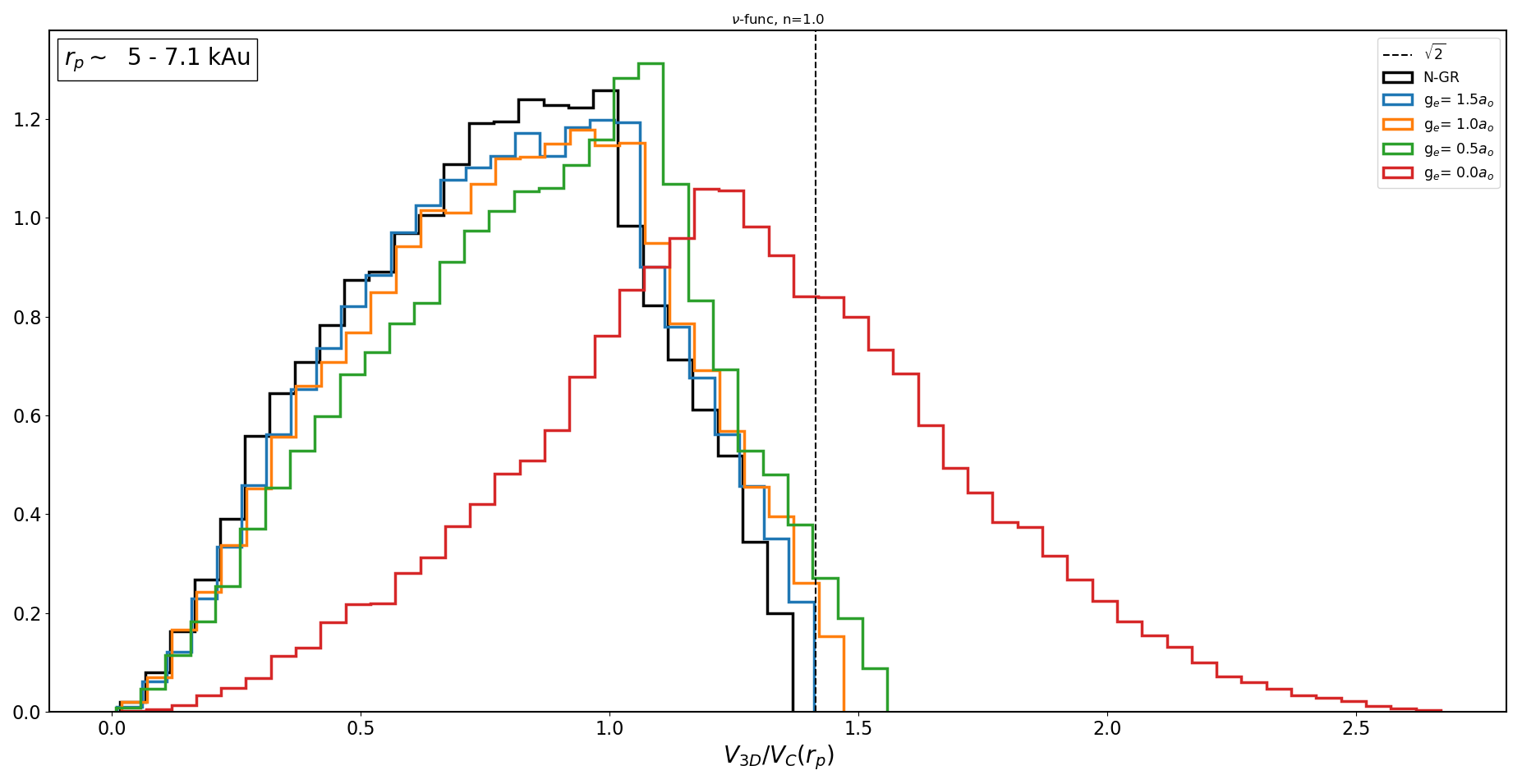} 
\caption{As Figure~\ref{fig:hist35-efe-simp} for projected
 separation range $r_p \in (5, 7.1) \kau$. 
 \label{fig:hist57-efe-simp} 
} 
 \includegraphics[width=13cm]{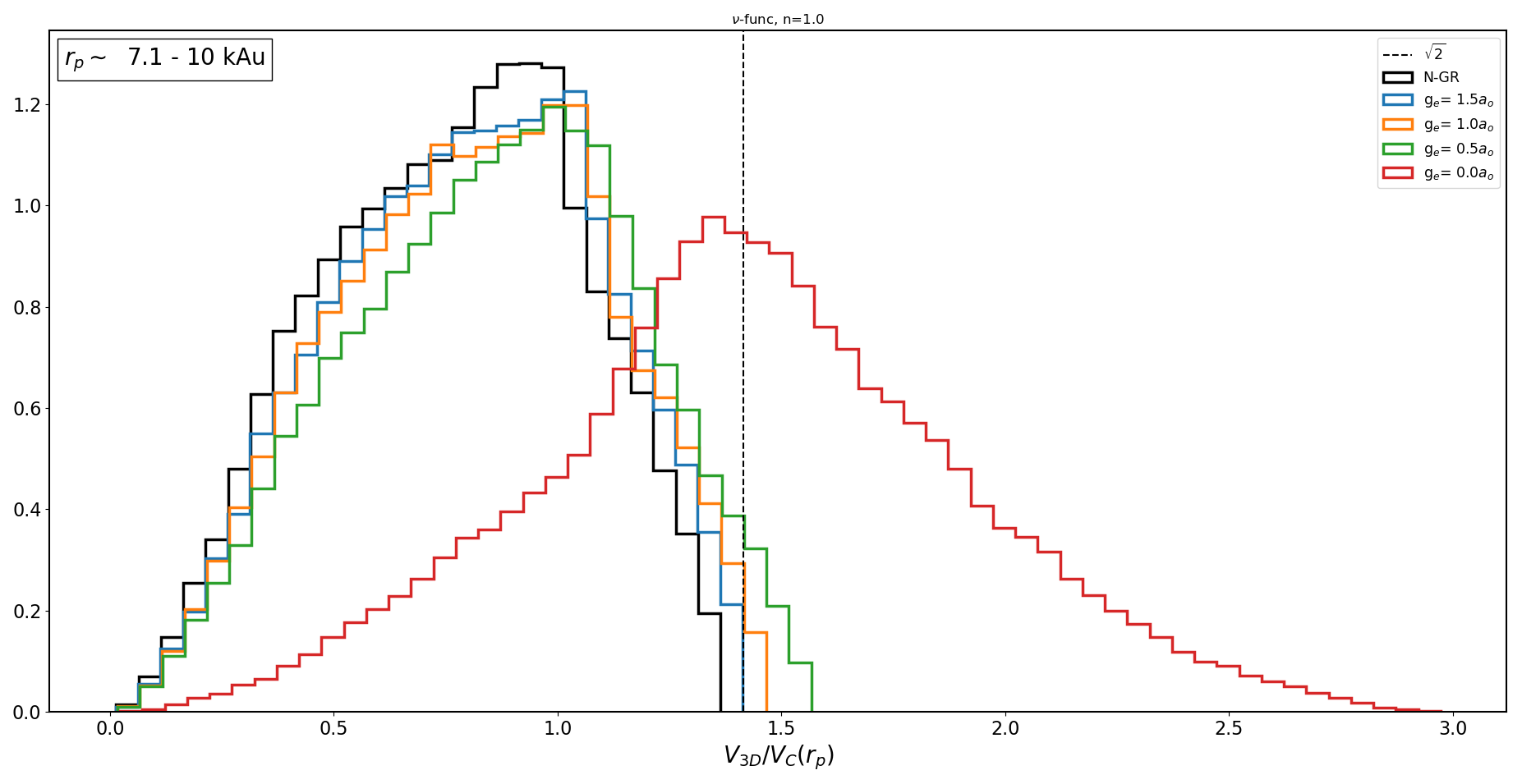} 
\caption{As Figure~\ref{fig:hist35-efe-simp} for projected
 separation range $r_p \in (7.1, 10) \kau$. 
 \label{fig:hist710-efe-simp} 
} 
\end{figure*}  

\begin{figure*} 
 \includegraphics[width=13cm]{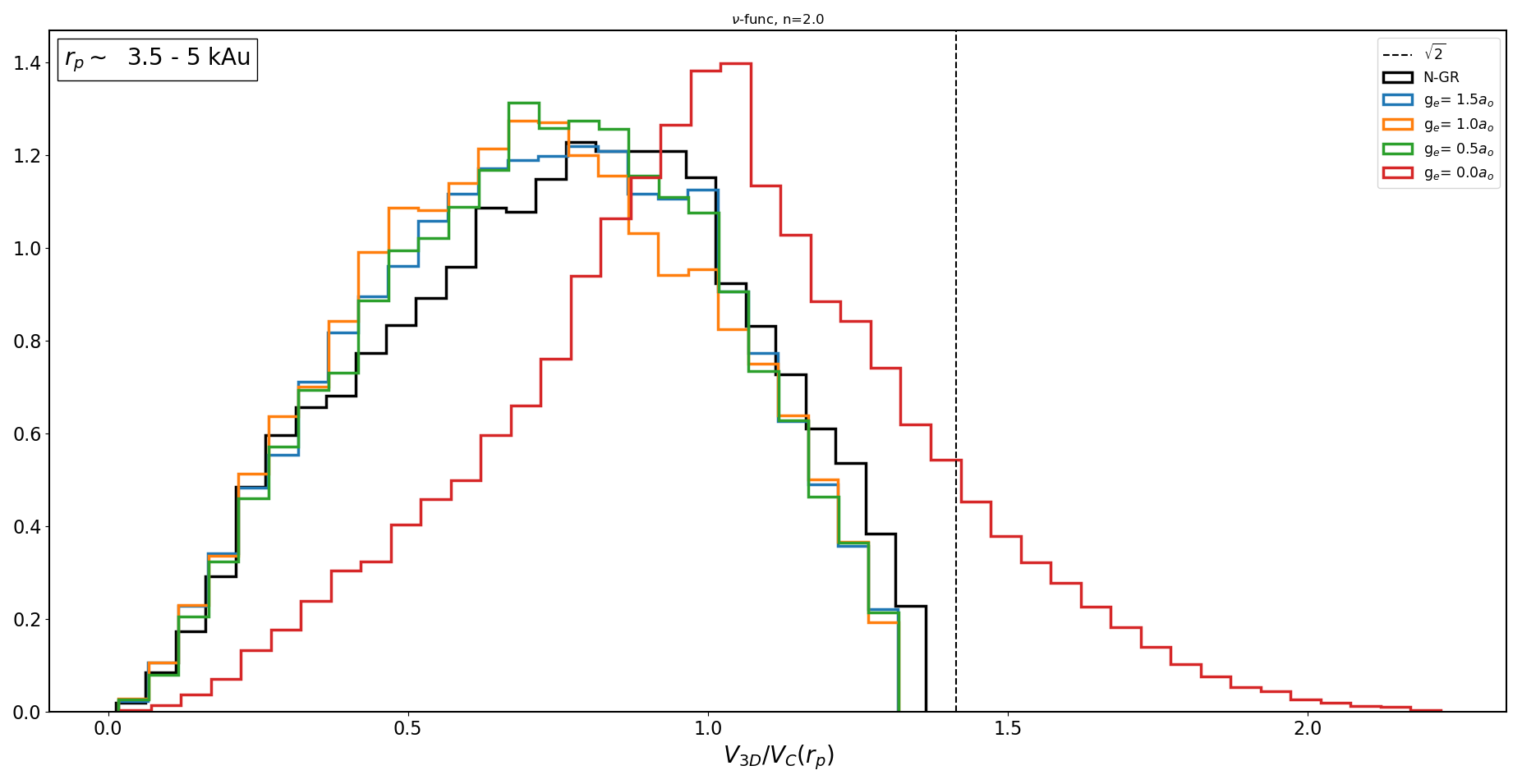} 
\caption{Histogram of simulated $\vthree/v_C(r_p)$ for MOND-standard
  interpolating function, with EFE, for
 projected separation range $r_p \in (3.5, 5) \kau$.  
 Black histogram is for standard Newtonian gravity, and 
 coloured histograms are for MOND standard interpolating function including 
  EFE with four values of external field $g_{Ne}$: $g_{Ne} = 0$ (red), 
  0.5 (green), 1.0 (yellow) and 1.5 (blue) in units of $a_0$.  
 \label{fig:hist35-efe-std} 
}
 \includegraphics[width=13cm]{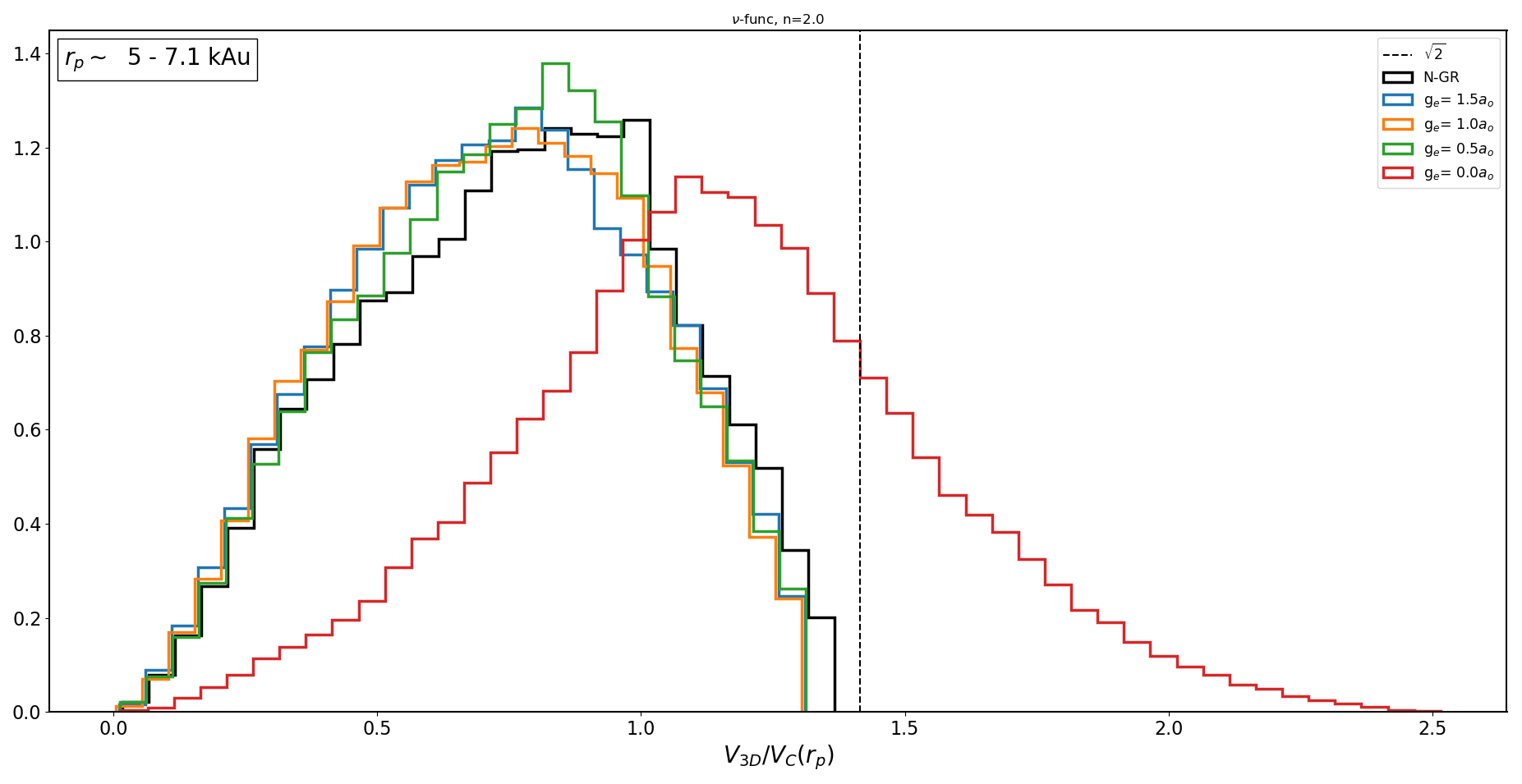} 
\caption{As Figure~\ref{fig:hist35-efe-std} for projected
 separation range $r_p \in (5, 7.1) \kau$. 
 \label{fig:hist57-efe-std} 
} 
 \includegraphics[width=13cm]{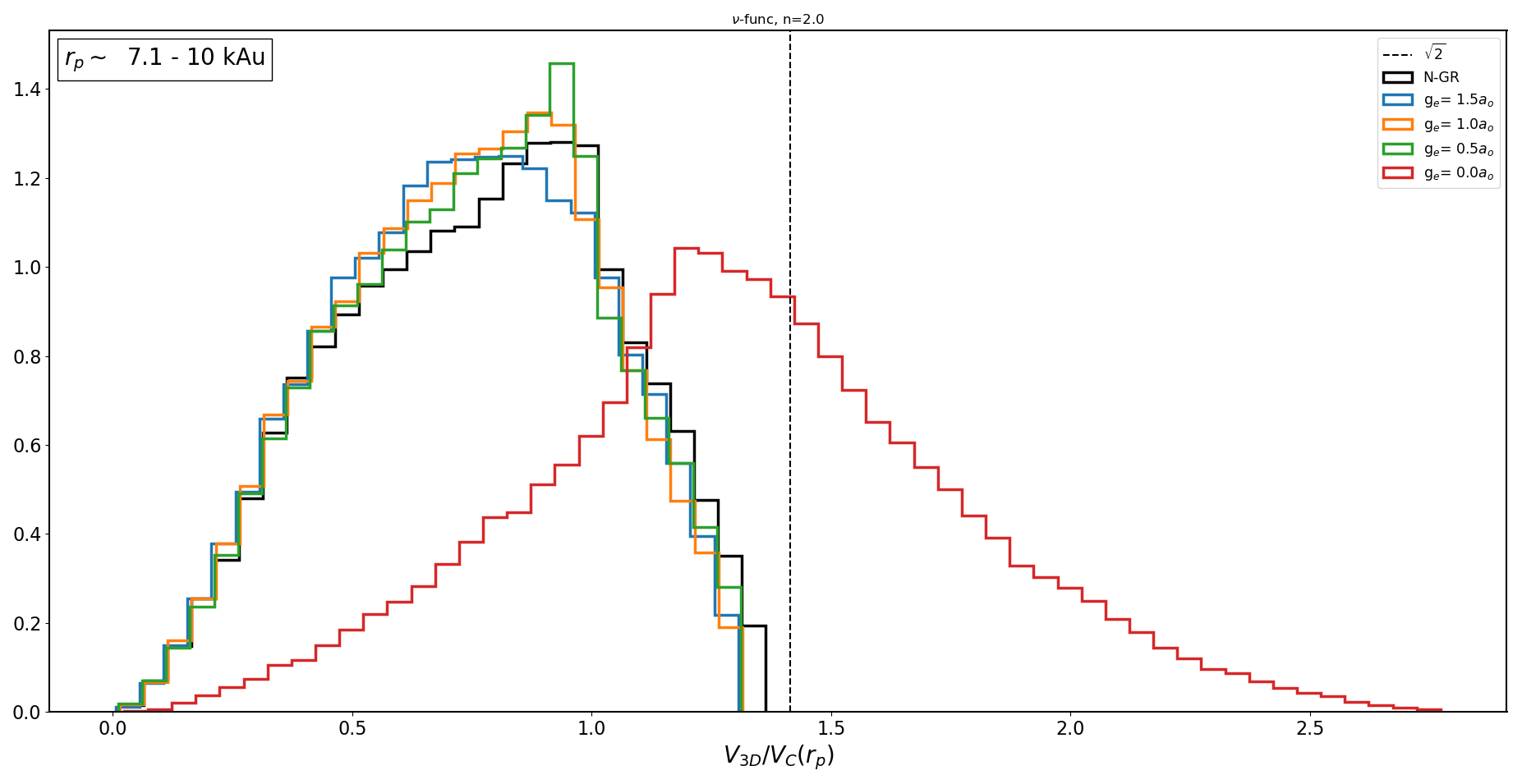} 
\caption{As Figure~\ref{fig:hist35-efe-std} for projected
 separation range $r_p \in (7.1, 10) \kau$. 
 \label{fig:hist710-efe-std} 
} 
\end{figure*}  

\begin{figure*} 
 \includegraphics[width=13cm]{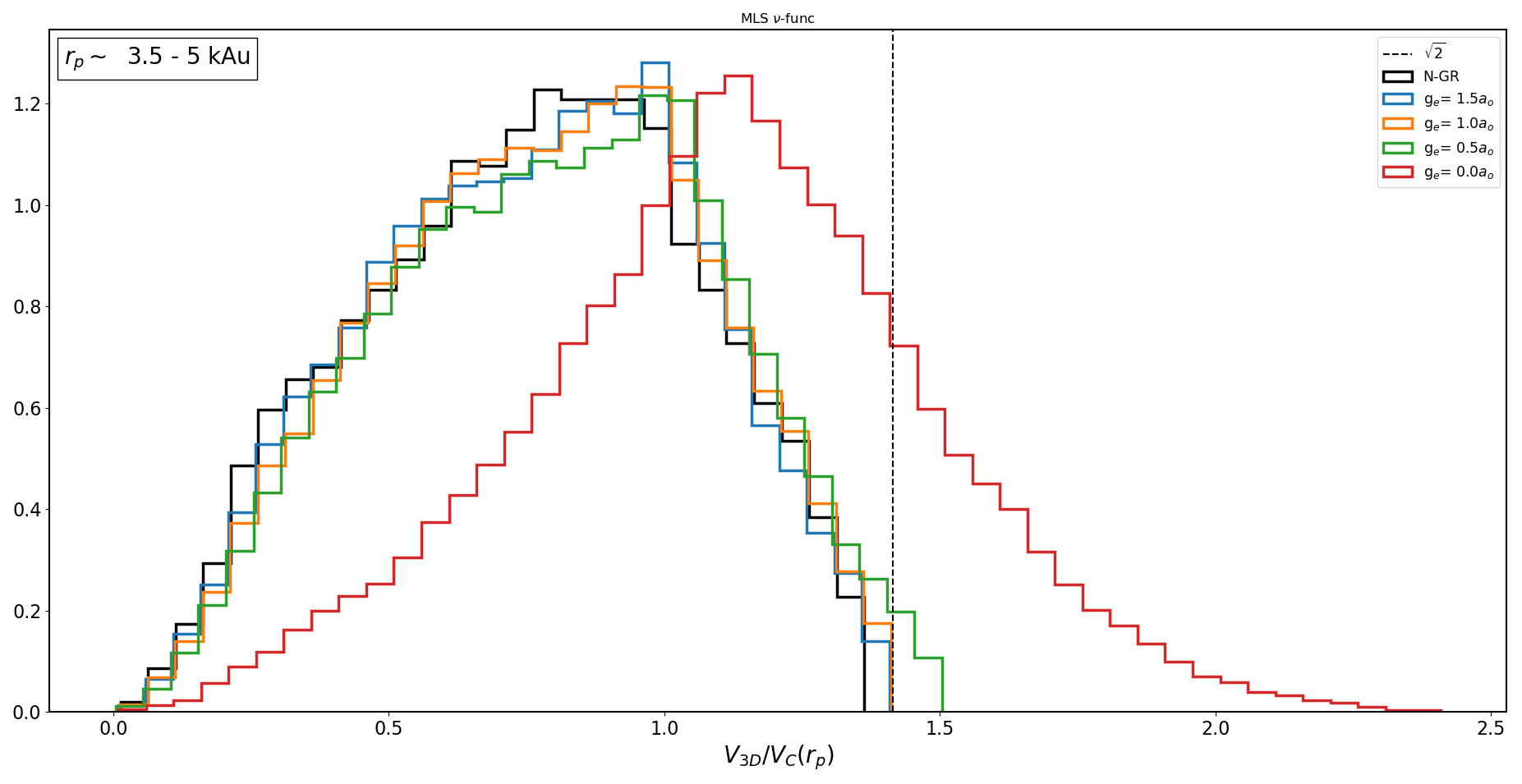} 
\caption{Histogram of simulated $\vthree/v_C(r_p)$ for MLS
  interpolating function, with EFE, for
 projected separation range $r_p \in (3.5, 5) \kau$.  
 Black histogram is for standard Newtonian gravity, and 
 coloured histograms are for MLS interpolating function including 
  EFE with four values of external field $g_{Ne}$: $g_{Ne} = 0$ (red), 
  0.5 (green), 1.0 (yellow) and 1.5 (blue) in units of $a_0$.  
 \label{fig:hist35-efe-mls} 
}
 \includegraphics[width=13cm]{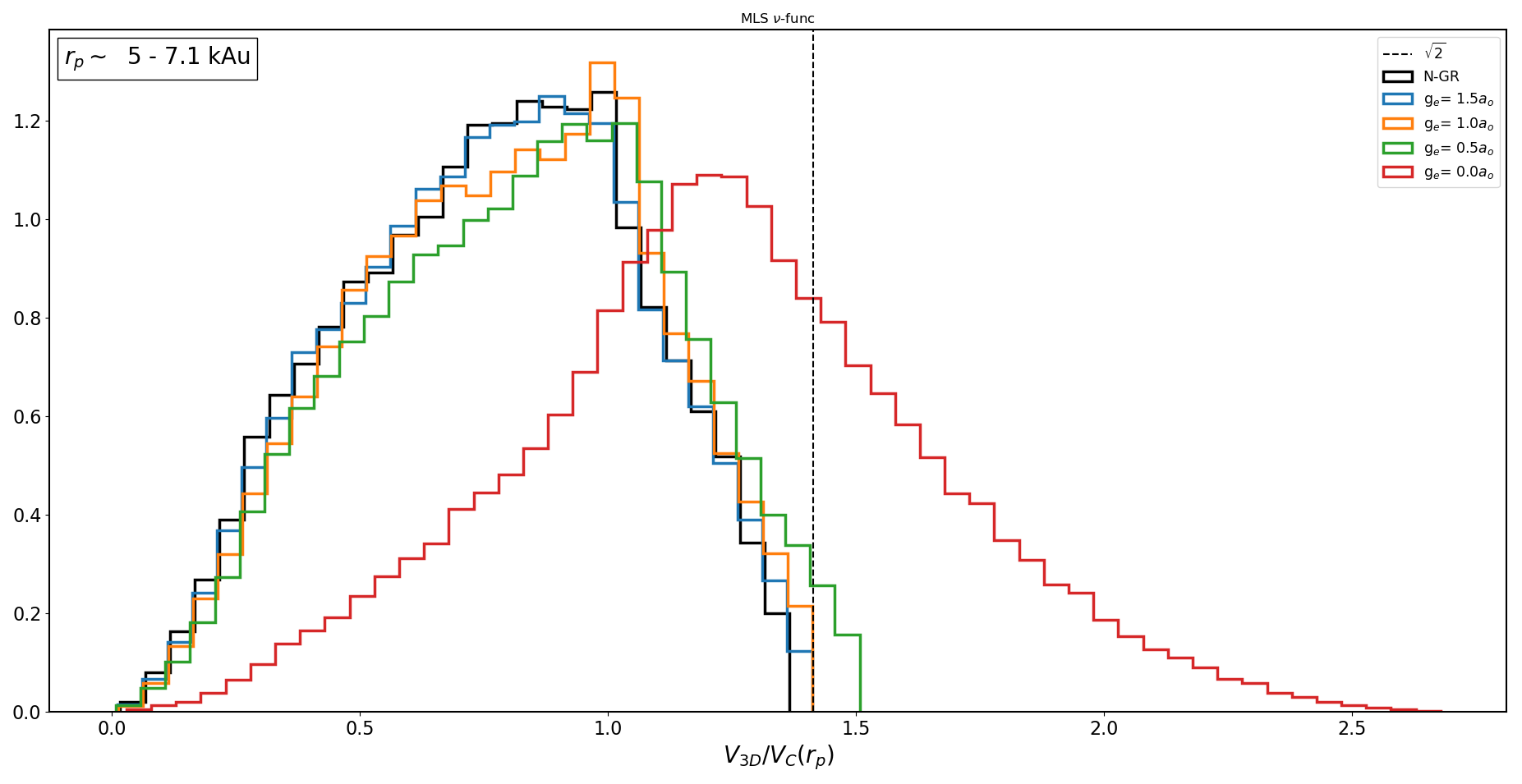} 
\caption{As Figure~\ref{fig:hist35-efe-mls} for projected
 separation range $r_p \in (5, 7.1) \kau$. 
 \label{fig:hist57-efe-mls} 
} 
 \includegraphics[width=13cm]{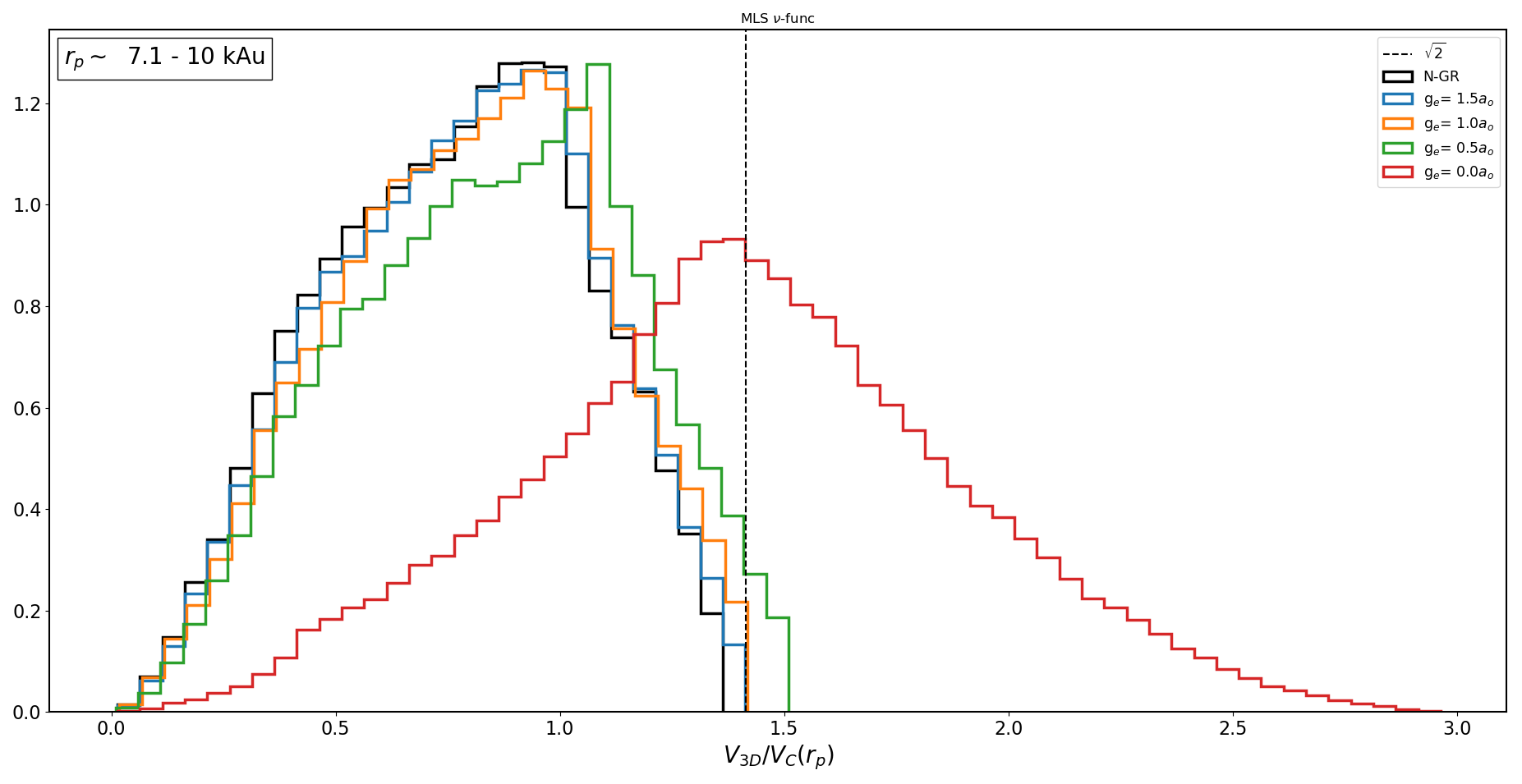} 
\caption{As Figure~\ref{fig:hist35-efe-mls} for projected
 separation range $r_p \in (7.1, 10) \kau$. 
 \label{fig:hist710-efe-mls} 
} 
\end{figure*}  


\begin{figure} 
\hspace*{-0.2cm}\includegraphics[width=9cm]{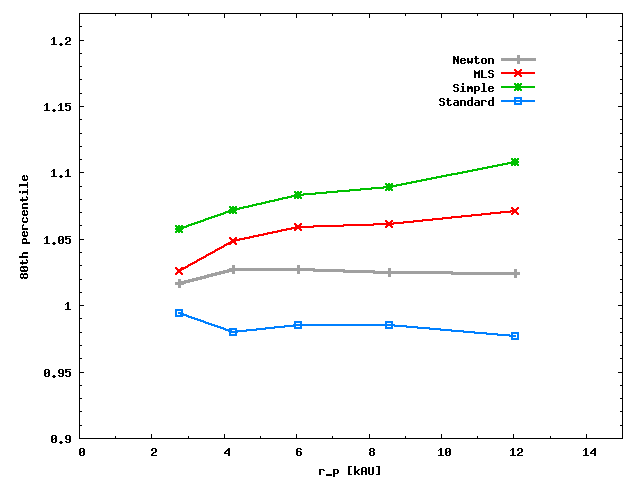} 
\vspace*{3mm} 
\hspace*{-0.2cm}\includegraphics[width=9cm]{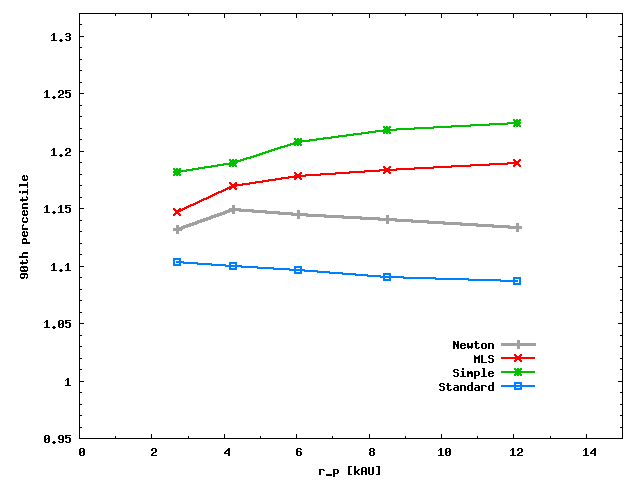} 
\caption{Upper panel: the 80th percentile values of $\vthree / v_C(r_p)$ 
  versus projected separation $r_p$ for Newtonian gravity, 
  and three MOND-like models including EFE with external 
  acceleration $G_{Ne} = 1 \, a_0$: the Simple interpolating
 function (topmost), MLS interpolating function (second from top) 
 and Standard interpolating
 function (bottom).  \newline 
 Lower panel: the same for the 90th percentile. \label{fig:perc80} 
}
\end{figure} 

The observed shifts in the relative-velocity percentiles (relative to 
 Newtonian values) are qualitatively as expected from the various
 MOND acceleration laws including the EFE, which behave
 roughly as a renormalisation of the gravitational constant 
   by a factor which is $\sim 0.9 - 1.3$ at low accelerations
 but converge back to 1 at $g_i \gg a_0$; this factor
 is generally quite slowly-varying over the range $0.3 \, a_0 < g_i < 2 \, a_0$
  of interest here, so the relative offset is only slowly 
  varying with binary projected separation and there is no sudden 
  feature at a specific projected separation. 

 We also find that exploring various choices of interpolation function 
 and different values of the ratio $g_e/a_0$, that the above rescaling
 factor is considerably sensitive to both the choice of interpolation function 
  and the numerical ratio of external acceleration $g_{Ne} / a_0$, 
  with the deviations increasing for smaller $g_{Ne}/a_0$.   
  We note that the TeVeS-like $\mu$ function (Eq.~\ref{eq:g_teves}) 
   produces relatively larger deviations than the others.

\subsection{The QUMOND formulation} 
{\newtwo
  After submission of the original version of this paper, 
  we became aware of the work of \citet{Banik15} and \citet{Banik18a}; 
 the latter concerns tidal streams rather than binaries, but 
 is also relevant to the case of wide binaries as follows.  
  The QUMOND formulation was introduced by \citet{Milgrom 2010} 
  as a simplification of the earlier aquadratic Lagrangian (AQUAL) 
  approach; using QUMOND, \citet{Banik15} 
  give a semi-analytic solution for the acceleration 
  due to a point mass embedded in a tidal field. 
  Banik (private communication) has supplied us with an example set
   of solutions for several MOND interpolating functions, and the
  result is that the deviations above Newtonian gravity are rather
   larger than those adopted above using approximation~(\ref{eq:g_efe}): 
  therefore,  a numerical application of 
   the QUMOND formulation to wide binaries 
   is likely to predict
  larger MOND effects and easier detectability 
    compared to the estimates here and below. Very recently a preprint has
  appeared by \citet{Banik18b}, with a rather detailed simulation  
   of wide binaries in QUMOND with the simple interpolating function; this
   indeed produces substantially larger MOND deviations than we 
  found here with approximation~(\ref{eq:g_efe}). }   

\section{Observational considerations}
\label{sec:obs} 

 We have seen above that {\em without} the EFE, we predict rather
 large and easily detectable shifts; however 
  with the EFE turned on (as generally expected), the shifts due to 
  modified-gravity are relatively small, so 
  it would clearly be necessary to obtain a rather large
 sample of wide-binary systems in order to get useful statistics. 
 We next make an approximate estimate of the number of useful 
  wide-binary systems as a function of limiting magnitude and distance, 
  to verify that \GAIA should produce a large enough sample that
 the purely statistical errors are small enough, subject to controlling all
  systematics. 

 For the present purposes we are mainly interested in binaries where
 both components are main-sequence stars of spectral type late-F, G, K 
  and early-M 
  (approximately $1.2 \msun$ to $0.5 \msun$) since 
  brighter stars have few metal lines 
 for precise RV's, while stars below about $0.5 \msun$ are intrinsically
  faint and hence observationally more challenging.  
 
\subsection{Number of   wide-binary systems}
\label{sec:num} 
Here we use the luminosity function (LF) 
 of \citet{Chabrier 2003} and the binary separation distribution 
 from \citet{Andrews 2017} to estimate the number 
 of suitable systems as a function of apparent magnitude, $V$, 
  distance to system, $d$, and separation of the stellar binary, $s$.\\
\\ 
The mass function of \citet{Chabrier 2003} 
 results in a local number density 0.0166 stars $\pc^{-3}$ with 
 $0.6 \le M/\msun \le 1.2$, a range well suited for precision 
 radial velocities. Assuming a 
  $300 \pc$ disk scale-height and adding a requirement $V < 15$ 
 gives an estimate of total $\sim 196,000$ such stars  
 within $D < 150 \pc$, or $\sim 394,000$ within 200 pc, 
   where the distance 
 threshold is chosen for good GAIA transverse velocity precision; 
 of course, only a small fraction of those will be members of wide-binary
 systems of interest here. 


The distribution in orbital separations has been estimated 
  recently by \citet{Andrews 2017}, who determined 
 that the distribution in projected separation $s$ is
  consistent with the traditional Opik law 
   $s^{-1}$ (i.e. flat in $\ln s$) up to $s \la 5 \kau$, with 
  evidence for a break to a steeper power-law $p(s) \propto s^{-1.6}$ at 
  separations $s \geq 5 \kau$.  
 If we assume that this broken power-law applies 
  from $0.03 \,{\rm AU}$ to $100 \kau$, and normalise so that 
   50 percent of FGK stars have a binary
  companion anywhere in that range, 
  this predicts that 5.5 percent of FGK stars have a companion star    
 in our range of interest $3 < s < 20 \kau$; the results
  of \citet{Lepine 2007} suggest a slightly higher fraction. 
   A significant minority of the secondaries will be mid-M or late-M stars 
  and thus too faint for practical RV followup,
 but we estimate that $\sim 3$ percent of the above FGK stars 
  should have a usable binary companion at $3 < s < 20 \kau$ and 
   $V \le 15$.   

  Combining the above leads to an approximate estimate of about 5000 
   potentially usable wide-binary systems (for $D < 150 \pc$, $V < 15$),  
   or over 10,000 if we extend to $D < 200\pc$; these are 
   divided between about five separation-bins as used above, 
  so around 1,000 to 2,000 systems per $\sqrt{2}$ separation bin.    
 
\subsection{Observational caveats} 
\label{sec:cav} 

 For binaries in our considered range $V \la 15$, $d \le 200 \pc$, 
  the \GAIA statistical proper-motion errors are $\la 0.02 \kms$, 
 and in principle modern planet-hunting spectrographs such 
 as HARPS and ESPRESSO can readily reach differential RV precision well below 
  $0.01 \kms$ at $V \sim 15$. Stellar RV jitter is also 
  usually negligible at these levels, so in principle the 
 binary relative velocities are measurable at better than the 10 percent 
  level.  

 However, other sources of  error are potentially more serious: 
   here we discuss some observational considerations regarding absolute
 velocity precision, contamination by unbound pairs, and 
  confusion from hierarchical triple/quadruple systems. 
Our method does rely on a rather good calibration of
 the luminosity-mass relation. However, this is potentially
 testable using the binaries at smaller projected separation $\la 1 \kau$
  where the deviations due to modified gravity are predicted
 to be nearly negligible. Also, the required high-quality spectra (for radial
 velocities) should provide precise metallicities, allowing this to 
  be included in the calibration.   

Concerning the absolute velocity precision, while modern high-stability 
 planet-search spectrographs can routinely deliver radial velocity precision 
  $\sim 0.001 \, {\rm km\, s^{-1}}$, this however is the differential precision
 over time for the {\em same} star; 
  while here we are interested in the {\em absolute} RV {difference} 
 between the two components of a wide binary.  This implies that unlike the 
  planet-finding case,  extra systematic effects
  such as gravitational redshift,   convective blueshift, 
 and zeropoint errors from spectral mismatch 
   (actually the difference in these between the two stars) 
   must be corrected for. 
 The gravitational redshift term is $\simeq 0.633 \kms (M/R)$ in solar
 units, and since the $M/R$ relation is fairly close to linear
  this is slowly varying for main-sequence stars.  This should probably be 
 correctable at better than the $0.03 \kms$ level.  

 The convective-blueshift term (arising from rising/falling
  cells of differing temperatures in the stellar atmosphere) 
 is somewhat more challenging,
  with amplitude estimated as $\simeq -0.3 \kms$ for the Sun and
 decreasing towards low-mass stars \citep{Kervella 2017}. 
 However in practice this term is calibrated out when RVs are zeropointed
   to the Sun, but re-appears for stars of non-Solar-like spectrum. 
 This term is probably the most important systematic effect 
  in limiting the absolute accuracy on radial velocities;   
 however, the potential bias can be reduced by either selecting subsamples
  of binaries of similar spectral type,  or tested by 
 studying the asymmetry of binary RV differences 
  vs spectral type, since the orbital velocity differences
  should on average be symmetrical around zero.  

 It is also possible to use short-period binaries 
  to test for systematic offsets in absolute
 radial velocity zero-point as a function of spectral type: 
  for short-period binary stars, the time-average of the two 
 radial velocities averaged over an integer multiple of the period 
  should both be equal to the barycentre radial velocity.  Thus, observations
 of medium-separation binary stars with known 
  moderate-period orbits and selected spectral types can potentially 
  provide a check for type-dependent shifts in the RV zeropoint. 

 Hierarchical triple/quadruple systems where one or both
 components of the wide system are themselves close binaries 
  are a more serious issue, 
  since for unequal masses (thus very unequal luminosities) 
 the extra components may lurk undetected and 
  can greatly shift the observed relative velocities of the wide 
    system compared to the value for an isolated binary.   
 However, we estimate that this source of contamination should usually be 
   removable by follow-up observations: 
  in the case of very close inner pairs $\la 3 \au$ 
  these should show large radial-velocity variations within a 
    timespan of a year or two; 
  while wider systems $\ga 3 \au$ 
  should be resolvable by direct imaging with adaptive optics unless
   the extra companion is very faint.  
 This leaves intrinsically faint brown-dwarf 
   or super-Jupiter companions with periods of order 10-100 years 
   as the main potential problem. 
  The brown-dwarf ``desert'' is helpful in this respect, as
  cold Jupiters only produce small reflex motions of order $0.012 \kms$,  
  while brown dwarfs above the D-burning limit should mostly be detectable 
   in deep imaging.  

 Contamination from unbound pairs misclassified as bound binaries 
   is a potentially more serious issue: 
  however, we note that for a random phase-space distribution the
  contamination should be very small.  For objects with a density 
 of order $0.1 \pc^{-3}$ and velocity dispersion $\sim 25 \kms$, 
 the probability for a given primary star to have a chance-flyby companion 
  with projected separation 
  $r_p < 30 \kau$, radial separation $\Delta d < 0.5 \pc$ and 
  3D velocity difference $\le 1 \kms$ is of order $10^{-6}$, which is far
   below the estimated fraction of true wide binaries.  
  This leaves the major issue as objects with correlations in phase-space: 
   either ``ionized'' previously-bound 
   wide binaries, or unbound pairs with a common origin, are the main
  potential source of contamination.   Most of these are expected to have 
  $v_{3D}/v_c(r_p) \ga 2$ and these can be clipped from the
  sample;   the remaining issue is that unbound common-origin  
   pairs aligned at a relatively
  small angle to the line-of-sight (small $\sin \beta$) can then 
   masquerade as bound binaries with $v_{3D}/v_c(r_p) \sim 1.1$, hence
  causing an upward bias in the 80th or 90th percentile value 
    for apparently-bound binaries.  
  It should be possible to model the distribution of these by
  counting unbound pairs as a function of projected separation 
  and velocity difference and assuming random viewing angles, though
   this will require further study; this is beyond the scope of the
  present work.   

 Thus it appears that none of the above problems is serious enough to 
  be a fundamental blocker from an observational perspective, 
  though they may substantially increase the requirements in 
  follow-up observing time 
 to eliminate hierarchical triple/quadruple systems, to check
  for spectral-type-dependent offsets in the RV zeropoint, 
  and to check the mass-luminosity relation using smaller-separation 
  binaries.  

\subsection{Statistical errors} 
\label{sec:stat} 
Here we note that the steep fall in the histogram of relative velocities
 at $v_{3D}/v_C(r_p) \sim 1.1$ is also helpful for statistics: this implies
 that the statistical uncertainty in estimating the 80th and 90th percentiles
 from a sample size of $N$ binaries  
 is substantially smaller than the naive 
  estimate $\approx 1 / \sqrt{N}$; in essence this arises because detecting a
 ``sharp edge'' in a distribution is more precise than 
  estimating the centroid of a broad distribution.  

For example, if we define $X_{90}$ to be the observed 90th percentile
 from a sample of $N$, and $x$ is an arbitrary variable, then
 $P(X_{90} > x)$ can be calculated as the binomial probability of obtaining 
  $\ge 0.1 N$ ``successes'' from $N$ 
  independent trials with probability $1-C(x)$, where 
  $C(x)$ is the cumulative PDF for one binary. 
 Then, for an example case of $N = 1000$, we expect $100 \pm \sqrt{90}$
   binaries above the {\em true} 90th percentile, hence there is just over
 68\% probability that the {\em observed} 90th percentile will fall between
  the true 89th and 91st percentiles; from the simulated histograms
  above, these points are offset by 
 $\simeq \pm 0.01$ from the 90th percentile.  Thus the uncertainty on the
 90th percentile should be reasonably approximated by $0.3/\sqrt{N}$, 
  not simply $1/\sqrt{N}$. 
 This implies that a sample of $\sim 1000$ well-measured binaries can give
 a statistically significant detections of an 
  offset $\sim 0.04$ relative to Newtonian predictions, 
 which is enough to robustly detect the offsets predicted in MOND-like
  modified gravity models,  even in the various EFE cases,
 {\em if} all systematic errors and contamination can be controlled
   well enough and/or statistically corrected via simulations.   

\section{Conclusions}
\label{sec:conc} 

{\newtwo Following on from earlier related 
 studies (e.g. \citet{Hernandez 2011}, 
 \citet{Hernandez 2012}, \citet{Hernandez 2017}), } 
 we have estimated the prospects for observational tests of 
  modified-gravity theories using wide-binary stars selected  
 by \GAIA and high-precision radial velocities from 
 ground-based telescopes.   Considering the ratio of
 3-D relative velocities to the Newtonian circular velocity, 
   in standard gravity the probability distribution function contains
  a rather steep decline at $v_{3D}/v_C(r) \sim 1.2$, resulting in 
  80th and 90th percentile values which are only
  weakly dependent on the uncertain distribution of orbital 
  eccentricities.  
 In a practical case we only have access to projected
  separation $r_p$ rather than $r$, 
  but this causes only a modest broadening of the distribution towards
  smaller values of the ratio $v_{3D}/v_C(r_p)$; this 
  parameter is well measurable
 with \GAIA data combined with accurate ground-based RV measurements.  

 We simulated large numbers of binary orbits with various gravity models
  observed at random angles and phases, and evaluated the
  statistical distribution of $v_{3D}/v_c(r_p)$, in particular
 the 80th and 90th percentile values which are reasonably insensitive 
  to the eccentricity distribution and only weakly
  sensitive to a small fraction of contaminants. 
 
 Our general conclusions are summarised as follows: 
\begin{enumerate} 
 \item If the relevant modified-gravity theory does {\em not} contain
  an external field effect,  then large non-Newtonian 
   deviations in the relative velocity
   distribution should be easily observable
  for binaries wider than about $5 \kau$, so MOND-like
  theories {\em without} an EFE should be rather  
   easy to detect or rule out with samples of a few hundred wide
  binaries. 
\item Binary projected separations of order $3 - 15 \kau$ 
   seem to be the most promising range, since MOND-like effects 
  should start to appear above a few $\kau$; 
  while other considerations (required
   velocity precision, perspective rotation effects, tidal effects)
   all become more challenging at even larger separations.  
\item With the external field effect (EFE) turned on  
   (as in most MOND-like modified gravity theories), the deviations 
  are considerably reduced, but are still potentially detectable
  and contribute a shift of order $\sim 4 - 8$ percent, depending
  on the MOND interpolating function,   which is 
   potentially detectable with a moderately large statistical 
   sample of order 1000 well-observed wide-binary systems. 
\item Again with the external field effect turned on, the size of deviations 
  predicted by MOND-like theories   are quite sensitive to 
  the specific shape of the MOND interpolating function (or equivalent) 
  and the value of the external field.   
  Since the Galactic acceleration field (from baryons) is 
   quite close to $1 \, a_0$,   
   the wide binaries are in a regime where the Galactic and internal
    accelerations are rather similar.  This implies that MOND-like
   effects tend to produce an acceleration law with slope fairly close to
   $1/r^2$, but with an apparent rescaling of the gravitational 
   constant.  
\item To a reasonable approximation, MOND-like theories
  including the EFE produce a shift in the 80th and 90th percentiles
  of $\vthree/v_C(r_p)$ 
    which are proportional to $\sqrt{\Geff /G}$ in 
   the relevant MOND model; this allows a qualitative assessment 
   of the effects for other MOND-like models beyond those simulated here.   
\item {\newtwo Improved constraints in future on the Galactic acceleration and 
  the required shape of the MOND interpolating function 
   around $\sim 1\, a_0$, both 
  from {\em GAIA} and external galaxy rotation curves, will be helpful
   to constrain these and give more specific predictions for the size
    of deviations}.    
\item  {\newtwo More detailed
  computations of the MOND accelerations with the Galactic external 
  field as in e.g. \citet{Banik15} and \citet{Banik18b} 
    indicate larger deviations
   from GR than the approximations we used above;  this improves
  the prospects for observational tests. }  
\item The sample of observable wide binaries at $d \la 200 \pc$
  from \GAIA is probably large enough to give a statistically significant 
  { \newtwo test for the presence or absence of } MOND-like effects, if all
  systematic errors and contamination can be well controlled or
   statistically corrected from simulations.  
\end{enumerate} 

 Further study is needed to investigate the practical effects of 
  contamination from common-origin unbound stellar pairs, realistic 
  systematic errors in the luminosity/mass relation, 
   possible radial velocity systematic errors, 
  and perspective-rotation effects; 
 these will probably require considerably more detailed simulations
  and also more input from future observations, so this paper 
  provides essentially an initial feasibility study.  

  However, this test looks potentially very interesting as 
   an observationally viable probe of possible 
   modified-gravity effects in a {\newtwo relatively less explored }  
  portion of parameter space.  

\section*{Acknowledgements}
 This is an author-produced, non-copy-edited version of the manuscript
  accepted by MNRAS.  The version of record is available at 
 Digital Object Identifier DOI:10.1093/mnras/sty1578 . 

 CP is supported by an STFC studentship. 
 We thank Tim Clifton and Indranil Banik 
  for helpful discussions. 
  We thank the referee, Xavier Hernandez, for 
  a helpful report and comments 
  which significantly improved this paper. 








\end{document}